\documentclass[fleqn,usenatbib]{mnras}

\usepackage{newtxtext,newtxmath}

\usepackage[T1]{fontenc}

\DeclareRobustCommand{\VAN}[3]{#2}
\let\VANthebibliography\thebibliography
\def\thebibliography{\DeclareRobustCommand{\VAN}[3]{##3}\VANthebibliography}

\usepackage{graphicx}	
\usepackage{amsmath}	

\usepackage{amssymb}	
\usepackage{threeparttable}

\newcommand*{\dif}{\mathop{}\!\mathrm{d}}

\title[SN 2019va]{SN 2019va: A Type IIP Supernova with Large Influence of Nickel-56 Decay on the Plateau-phase Light Curve}

\author[X. Zhang et al.]{
Xinghan Zhang$^{1}$,\thanks{E-mail: xh-zhang17@mails.tsinghua.edu.cn},
Xiaofeng Wang$^{1,2}$\thanks{E-mail: wang\_xf@mail.tsinghua.edu.cn},
Hanna Sai$^{1}$,
Jun Mo$^{1}$,
A. P. Nagy$^{3}$,
Jicheng Zhang$^{4}$,
Yongzhi Cai$^{1}$,
\newauthor
Han Lin$^{1}$,
Jujia Zhang$^{5,6,7}$,
E. Baron$^{8}$,
J. M. DerKacy$^{8}$,
T.-M. Zhang$^{9,10}$,
Zhitong Li$^{9,10}$,
Melissa Graham$^{11}$,
\newauthor
F. Huang$^{12}$
\\
$^{1}$Physics Department and Tsinghua Centre for Astrophysics (THCA), Tsinghua University, Beijing 100084, China\\
$^{2}$Bejing Planetarium, Beijing Academy of Science and Technology, Beijing 100044, China\\
$^{3}$Department of Optics and Quantum Electronics, University of Szeged, 6720 Szeged, Hungary\\
$^{4}$Department of Astronomy, Beijing Normal University, Beijing, 100875, China\\
$^{5}$Yunnan Observatories, Chinese Academy of Sciences, Kunming 650216, China \\
$^{6}$Key Laboratory for the Structure and Evolution of Celestial Objects, Chinese Academy of Sciences, Kunming 650216, China\\
$^{7}$Center for Astronomical Mega-Science, Chinese Academy of Sciences, 20A Datun Road, Chaoyang District, Beijing 100012, China\\
$^{8}$Homer L. Dodge Department of Physics and Astronomy, University of Oklahoma, Norman, OK 73019, USA\\
$^{9}$Key Laboratory of Optical Astronomy, National Astronomical Observatories, Chinese Academy of Sciences, Beijing 10101, China\\
$^{10}$School of Astronomy and Space Science, University of Chinese Academy of Sciences, Beijing 101408, China\\
$^{11}$DiRAC Institute, Department of Astronomy, University of Washington, Box 351580, U.W., Seattle WA 98195, USA\\
$^{12}$Department of Astronomy, Shanghai Jiao Tong University, Shanghai, 200240, China
}

\date{Accepted XXX. Received YYY; in original form ZZZ}

\pubyear{2021}

\begin{document}
\label{firstpage}
\pagerange{\pageref{firstpage}--\pageref{lastpage}}
\maketitle

\begin{abstract}
We present multi-band photometric and spectroscopic observations of the type II supernova, (SN) 2019va, which shows an unusually flat plateau-phase evolution in its $V$-band light curve. Its pseudo-bolometric light curve even shows a weak brightening towards the end of the plateau phase. These uncommon features are related to the influence of $^{56}$Ni decay on the light curve during the plateau phase, when the SN emission is usually dominated by cooling of the envelope. The inferred $^{56}$Ni mass of SN 2019va is 0.088$\pm$0.018\,$M_{\odot}$, which is significantly larger than most SNe II. To estimate the influence of $^{56}$Ni decay on the plateau-phase light curve, we calculate the ratio (dubbed as $\eta_{\rm Ni}$) between the integrated time-weighted energy from $^{56}$Ni decay and that from envelope cooling within the plateau phase, obtaining a value of 0.8 for SN 2019va, which is the second largest value among SNe II that have been measured. After removing the influence of $^{56}$Ni decay on the plateau-phase light curve, we found that the progenitor/explosion parameters derived for SN 2019va are more reasonable. In addition, SN 2019va is found to have weaker metal lines in its spectra compared to other SNe IIP at similar epochs, implying a low-metallicity progenitor, which is consistent with the metal-poor environment inferred from the host-galaxy spectrum. We further discuss the possible reasons that might lead to SN 2019va-like events.

\end{abstract}




\section{Introduction}
\label{sec:intro}

Type II supernovae (SNe II) are believed to originate from the core collapse of massive stars with initial masses larger than 8\,$M_{\odot}$ \citep{Heger2003}. They are characterised by prominent P-Cygni profiles of Balmer series in their spectra \citep{Filippenko1997}. Photometrically, they are classified as SNe IIP if their light curves show extended plateau features ($\sim$100\,days), and SNe IIL if their light curves display post-peak linear declines \citep{Barbon1979}. Such a two-type classification is favoured by some statistical studies of SNe II (e.g., \citealt{Faran2014a, Faran2014b}). However, \citet{Anderson2014} pointed out that if the sample is sufficiently large, one will find that SNe IIP and SNe IIL belong to a continuous distribution. Such a distribution is believed to be related to the mass of the hydrogen envelope maintained by the progenitor stars, as evidenced by the results that SNe having larger envelope masses produce light curves with shallower plateau slopes and longer plateau durations \citep{Barbon1979, Swartz1991, Blinnikov1993}. In spite of the variety of envelope masses, the initial masses of the progenitor stars of SNe II are restricted to a certain range. Stellar evolution theory predicts an upper limit of 25\,$M_\odot$ for SN-II progenitors \citep{Heger2003}. Based on analyses of pre-explosion images (e.g. \citealt{Smartt2009}), the progenitors of most SNe II are found to be red supergiants (RSGs), but their masses lie in a narrow range (i.e., $\sim9-17$\,$M_{\odot}$). This inconsistency is known as the ``RSG problem'' and can be partially explained by the ``failed SNe'' theory \citep{Lovegrove2013}. As SNe II can also be used to determine distances, thus they are intriguing to the cosmology community. The measurement methods include the Expanding Photosphere Method (EPM, \citealt{Kirshner1974, Hamuy2001PhDT}) and the Standard Candle Method (SCM, \citealt{Hamuy2002Pinto}). The basic idea of EPM is to derive the intrinsic luminosity from the photospheric radius and the temperature, and then compare the intrinsic luminosity with the apparent value to obtain the distance, while the SCM is based on the correlation between the expansion velocity and the luminosity at a specific epoch.\\

There are many well studied SNe II in the literature, including SN 1999em \citep{Leonard2002, Elmhamdi2003}, SN 2005cs \citep{Pastorello2006, Pastorello2009}, SN 2013ej \citep{Valenti2014, Huangfang2015}, and SN 2017eaw \citep{Tsvetkov2018, Rui2019, VanDyk2019, Szalai2019, Buta2019, Szalai2019}. Light curves of these SNe II can be roughly divided into four phases: an initial fast rise, a plateau phase, followed by a transition to the final tail phase. This light-curve behaviour can be reproduced by several explosion models with different levels of simplifications (e.g., \citealt{Litvinova1985}, \citealt{Kasen2009}, \citealt{Pumo2011}, \citealt{Morozova2015}, \citealt{Nagy2014, Nagy2016}). \\

Early studies of plateau-phase light curves of SNe II only included the thermal energy deposited in the envelope by the explosion shock (e.g. \citealt{Litvinova1985}; \citealt{Popov1993}), but later studies revealed that the contribution of $^{56}$Ni decay during the plateau phase needs to be considered (e.g. \citealt{Kasen2009}; \citealt{Bersten2011}). \citet{Nakar2016} found that the influence of $^{56}$Ni decay on the plateau-phase light curve is common for SNe II. \citet{Kozyreva2019} further investigated the details of this influence by simulations, finding that $^{56}$Ni decay can extend and/or flatten the light curves of SNe II during the plateau phase, depending on the level of $^{56}$Ni mixing in the envelope. They also note that deducing the properties of the progenitor from the plateau-phase light curve may not produce accurate estimates if we only consider the thermal energy deposited by the shock. On the other hand, differences in the hydrogen-envelope mass have been proposed to account for the observed diversities in the statistical properties of SNe II. For instance, SNe II with less massive hydrogen envelopes tend to show more luminous $V$-band maximum ($M_{\rm max}^V$) and larger decline rates during the plateau phase ($s_2$) \citep{Anderson2014}. However, these correlations have large scatter, which may be reduced after including the influence of $^{56}$Ni decay. Quantifying such influences can also help better standardize the luminosity of SNe II and hence improve their use as distance indicators. SN 2019va is a type II supernova, for which $^{56}$Ni decay likely has a large influence on the plateau-phase light curve. The study of this SN can help further demonstrate the role of $^{56}$Ni decay in shaping the light curves of SNe II.\\

To contextualize SN 2019va within the SNe II landscape, we utilize several samples from the literature. For instance, we include the sample from \citet{Nakar2016} when discussing the influence of $^{56}$Ni decay on the plateau-phase light curve. This sample contains 24 SNe II that have enough multi-band photometric data to allow quantitative analysis of this influence. We supplement the \citet{Nakar2016} sample with 14 additional SNe II (see Table \ref{tab:etaNi} and Table \ref{tab:info}) from the literature that can have similar analysis. From this "24+14" sample, we select six representative, well-studied SNe II for the following comparison studies of light curves and spectra, i.e., SN 1999em \citep{Leonard2002, Elmhamdi2003}, SN 2005cs \citep{Pastorello2006, Pastorello2009}, SN 2009ib \citep{Takats2015}, SN 2013by \citep{Valenti2015}, SN 2016gfy \citep{Singh2019}, and SN 2017eaw \citep{Tsvetkov2018, Rui2019, VanDyk2019, Szalai2019}. Additionally, the SN-II samples collected by \citet{Anderson2014} and \citet{Valenti2016} are also used when discussing photometric correlations during the plateau phase.

This work is orgnized as follows: in Section \ref{sec:obs}, we describe the observations and the data reduction. In Section \ref{sec:host}, we analyse the properties of the host galaxy of SN 2019va. We present the photometric and spectroscopic evolution of SN 2019va in Section \ref{sec:phot} and Section \ref{sec:spec}, respectively. In Section \ref{sec:discuss}, we estimate the synthesised $^{56}$Ni mass in SN 2019va, and discuss the influence of $^{56}$Ni decay on the plateau-phase light curve and hence on the explosion parameters. A summary is given in Section \ref{sec:conclusion}.\\

\section{Observations and data reduction}
\label{sec:obs}

\begin{figure*}
	\includegraphics[width=2\columnwidth]{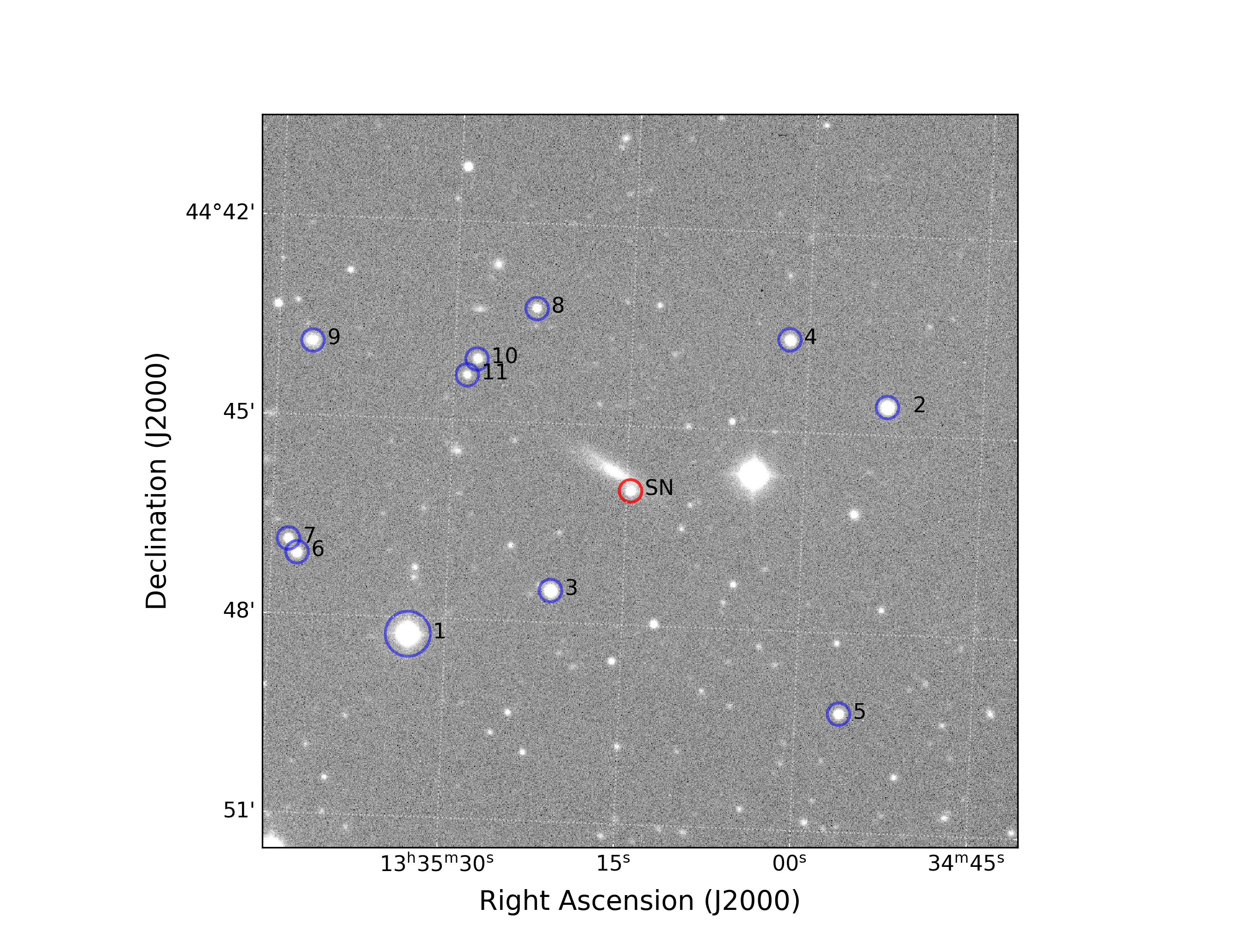}
    \caption{Finder chart of SN 2019va in galaxy UGC 08577. This image was in $r$-band and taken by the 0.8m Tsinghua-NAOC telescope on Feb. 1st 2019. SN 2019va is marked by a red circle and the local reference stars are marked by blue circles. See Table \ref{tab:refstars} for more information about the reference stars.}
    \label{fig:image}
\end{figure*}

SN 2019va was first detected on 2019-01-15.7 (UT dates are used throughout the paper; MJD=58498.7) by the Asteroid Terrestrial-impact
Last Alert System (ATLAS; \citealt{Tonry2018, Tonry2019}). It was located at $\alpha$=13:35:14.680 and $\delta$=+44:45:58.64, with offsets of $17.63\arcsec$ north and $16.24\arcsec$ west from the centre of the host galaxy, UGC 08577 \citep{Tonry2019}, as shown in Figure \ref{fig:image}. Three days after detection, a spectrum of SN 2019va was taken by \citet{Zhangjujia2019}, which showed a blue continuum with prominent P-Cygni profiles of the Balmer series, consistent with that of a type II supernova.\\

\subsection{Photometry}
\label{subsec:phot}
The $BVgri$-band photometric observations of SN 2019va started on Jan. 21st 2019 by the 0.8m Tsinghua-NAOC telescope (TNT hereafter) at Xinglong Observatory in China \citep{Huang2012}. All the obtained images were preprocessed with standard routines, including bias correction, flat fielding, and removal of cosmic rays. Then the images were reduced with a customized pipeline ``zrutyphot'' (J. Mo, in prep.). A template was first subtracted from the images and PSF photometry was performed on the differential images. The instrumental magnitudes were converted into the standard Johnson $BV$ and SDSS $gri$ system based on a series of local stars from the APASS \citep{Henden2009, Henden2010} and Pan-STARRS catalogues \citep{Magnier2020}. The final results are listed in Table \ref{tab:TNTphot}.\\

To obtain additional photometric data, we include the $c$- and $o$-band data of SN 2019va from the ATLAS Forced Photometry server\footnote{\url{https://fallingstar-data.com/forcedphot/}} \citep{Tonry2018, Smith2020}. The server calculates the flux at the SN location through ``forced photometry'', where a point spread function (PSF) is first estimated based on the nearby bright stars, and then a forced fitting of the PSF profile is performed at the SN position. For those nights when more than one measurement is available from ATLAS, we adopt the average flux for each night, weighted by the reciprocal of the squared error. The fluxes are then converted into magnitudes. The earliest ATLAS point was taken on MJD=58494.67 in the $o$ band with a flux value of $61.00\pm48.00~\mu$Jy, where 48.00 represents the 1-$\sigma$ uncertainty (Note that the ATLAS only took one data point on this night, so this value is not a weighted average). Based on a $3\sigma$ criterion, this measurement should be considered as an upper limit, i.e. $<3\times 48~\mu$Jy, which corresponds to >18.50\,mag in the $o$ band. Thus the next $o$-band measurement taken on MJD=58498.67 with a value of 16.91$\pm$0.03\,mag represents the first valid detection of SN 2019va. The photometric results from ATLAS are presented in Tables \ref{tab:ATophot} - \ref{tab:ATcphot}.\\

\begin{figure*}
	\includegraphics[width=1.8\columnwidth]{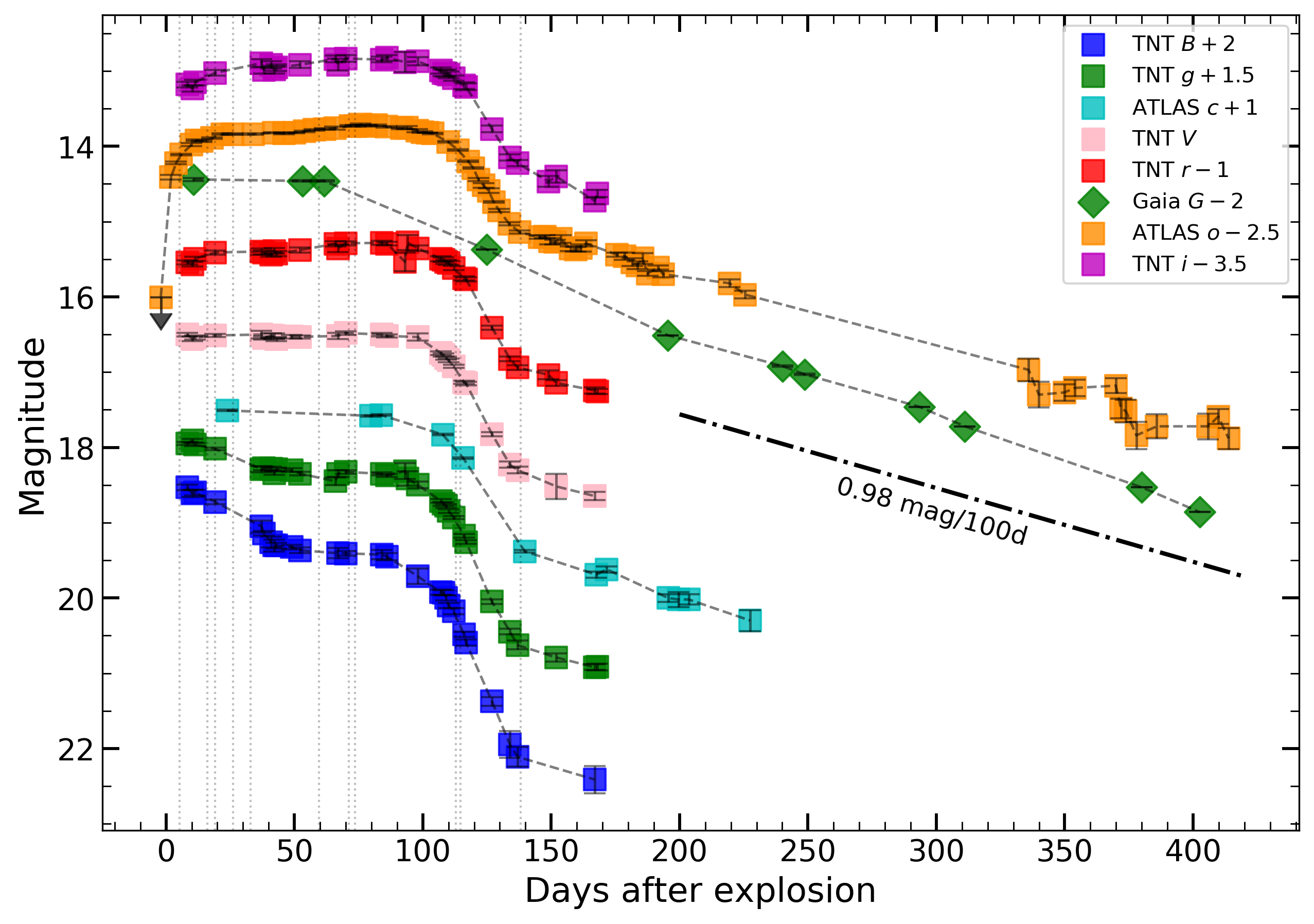}
    \caption{Multi-band light curves of SN 2019va. Data points are shown as squares and linked by dashed lines. The upper limit is denoted by a downward arrow. Those vertical dotted lines denote spectral epochs. The dot-dashed line with a slope of 0.98\,mag\,(100d)$^{-1}$ indicates the decay rate of $^{56}$Co $\rightarrow$ $^{56}$Fe.}
    \label{fig:LC}
\end{figure*}

In addition, SN 2019va was also monitored by Gaia for $\sim 400$ days, though with a low cadence. We download the data from ``Gaia Photometric Science Alerts'' (GPSA\footnote{\url{http://gsaweb.ast.cam.ac.uk/alerts/alert/Gaia19aht/}}). The photometric results are listed in Table \ref{tab:Gaiaphot}. All the photometric data are shown in Figure \ref{fig:LC}.

\subsection{Spectroscopy}

We obtained a series of spectra for SN 2019va using the 2.16-m telescope at Xinglong Observatory in China (XLT hereafter; \citealt{Jiangxiaojun1999, Zhangjicheng2016, Fanzhou2016}), the 2.40m telescope at Lijiang Observatory in China (LJT hereafter; \citealt{Chendong2001, Wangchuanjun2019}), and the 3.5m telescope at the Apache Point Observatory (APO hereafter). The spectra were reduced using the IRAF\footnote{\textsc{iraf} is distributed by the National Optical Astronomy Observatories, which are operated by the Association of Universities for Research in Astronomy, Inc., under cooperative agreement with the National Science Foundation (NSF)} routines, including bias correction, flat fielding, and removal of cosmic rays. The wavelength was calibrated through comparison with the arc lamp spectra, and the flux was calibrated using standard stars observed at similar airmass on the same night. The spectra were further corrected for atmospheric extinction, and telluric lines were removed from the data.\\

Two additional spectra are available from the TNS\footnote{\url{https://www.wis-tns.org//object/2019va}} website. One was taken on Mar. 14th 2019 by the Zwicky Transient Facility (ZTF, \citealt{Bellm2017, Bellm2019, GrahamM2019}) team, and the other was taken on Mar. 28th 2019 by the APO 3.5-m telescope.\\

\begin{figure}
	\includegraphics[width=\columnwidth]{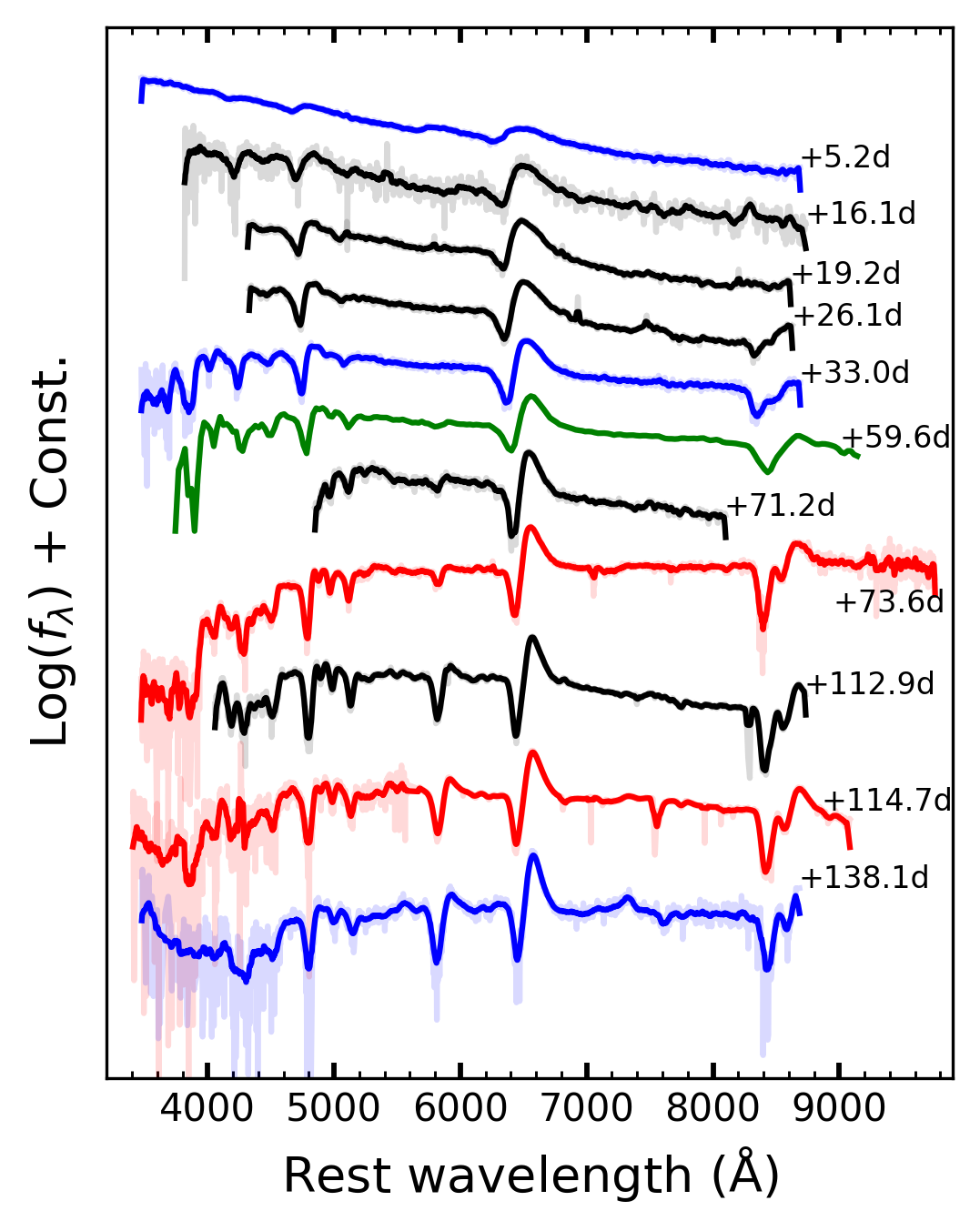}
    \caption{Spectral evolution of SN 2019va. Spectra from different instruments are indicated by different colours, with blue for the LJT, black for the XLT, red for the APO, and green for the ZTF, respectively. Phases marked near the spectra are relative to the explosion date, MJD=58496.7.}
    \label{fig:spectra}
\end{figure}

A journal of these spectra is presented in Table \ref{tab:spectra}, and they are shown in Figure \ref{fig:spectra}.

\section{Properties of the host galaxy}
\label{sec:host}

The host galaxy of SN 2019va, UGC 08577, is an edge-on spiral galaxy. Assuming H$_0$ = 73\,km\,s$^{-1}$\,Mpc$^{-1}$, $\rm \Omega_{\rm M}=0.27$, and $\rm \Omega_{\rm \Lambda}=0.73$, a distance modulus of $\mu=33.20\pm0.15$\,mag can be obtained from the NASA/IPAC Extragalactic Database. The Milky-Way reddening in the direction of SN 2019va is $E(B-V)_{\rm MW}=0.016$\,mag \citep{Schlafly2011}. To estimate the extinction from the host galaxy, we inspected the spectra of SN 2019va carefully, but we did not find any prominent \ion{Na}{i~D} absorption due to the host galaxy. Therefore, we adopt the Milky-Way extinction as the total extinction, i.e., $E(B-V)_{\rm total}=0.016$\,mag.\\

To determine the star formation rate (SFR) of UGC 08577, we perform photometry on this galaxy with an elliptical aperture in a background-subtracted \textit{GALAX}-FUV-band image\footnote{downloaded from the MAST website (\url{https://mast.stsci.edu/portal/Mashup/Clients/Mast/Portal.html})} and obtain $\sim$17.68\,mag. As SN 2019va is located at the outskirts of the host galaxy, the reddening derived above thus cannot represent the full-galaxy reddening. However, a spectrum near the centre of UGC 08577 was taken during the Sloan Digital Sky Survey (SDSS; \citealt{Adelman-McCarthy2008}), from which the flux ratio of H$\alpha$/H$\beta$ is measured as $\sim$3.954. According to Equation (4) of \citet{Dominguez2013}, such a value corresponds to a reddening of $E(B-V)\approx 0.277$\,mag. Since the galaxy centre is the most luminous part, we adopt this value as the full galaxy reddening. Assuming the \citet{Fitzpatrick1999} extinction law, we obtain the extinction-corrected FUV magnitude for the host galaxy as $\sim$15.52\,mag. Based on Equation (3) of \citet{Karachentsev2013}, this FUV magnitude gives a SFR of $\sim$0.714\,$M_{\odot} \rm yr^{-1}$, corresponding to a birth rate of one SN II per $\sim$156 years in UGC 08577 according to the relation between SFR and the core collapse SNe rate \citep{Singh2019}.\\

Line flux ratios inferred from host-galaxy spectra can be further used to estimate the metallicity of the galaxy. The logarithm (base 10) of the flux ratio between [\ion{N}{ii}] $\lambda6584$ and H$\alpha$ is called N2, while that between [\ion{O}{iii}] $\lambda5007$ and H$\beta$ is called O3. The value of N2-O3 is denoted as N2O3. After calculating the N2O3 value and substituting it into Equation (2) of \citet{Marino2013}, we obtain an oxygen abundance of $12+\log_{10}(\rm O/H)$= 8.36$\pm$0.03 for UGC 08577. This corresponds to a sub-solar metallicity (0.46$\pm$0.06$Z_{\odot}$) if we adopt the solar oxygen abundance as 8.69$\pm$0.05 \citep{Asplund2009}. Note that this value represents the metallicity near the centre of UGC 08577, while SN 2019va lies at the edge of this galaxy. It is well known that the abundance of heavy elements decreases with the distance from the galactic centre \citep{Henry1999, Sanchez2014}. Therefore, we adopt this value as an upper limit of the metallicity of SN 2019va (see also the discussion in Section \ref{subsec:pEW_FeII}).

\section{Photometric analysis}
\label{sec:phot}

\subsection{Estimate of the explosion epoch}

The explosion epoch can be constrained by the last nondetection and the first detection. We denote the former epoch as $t_{\rm ln}$ (i.e., MJD=58494.7 in the $o$ band) and the latter as $t_{\rm fd}$ (i.e., MJD=58498.7 in the $o$ band). Then the explosion epoch, $t_0$, is calculated as $(t_{\rm ln}+t_{\rm fd})/2$, and the error is estimated as $(t_{\rm fd}-t_{\rm ln})/2$. With this method, we estimate the explosion epoch as MJD=58496.7$\pm$2.0. \\

\begin{figure}
	\includegraphics[width=\columnwidth]{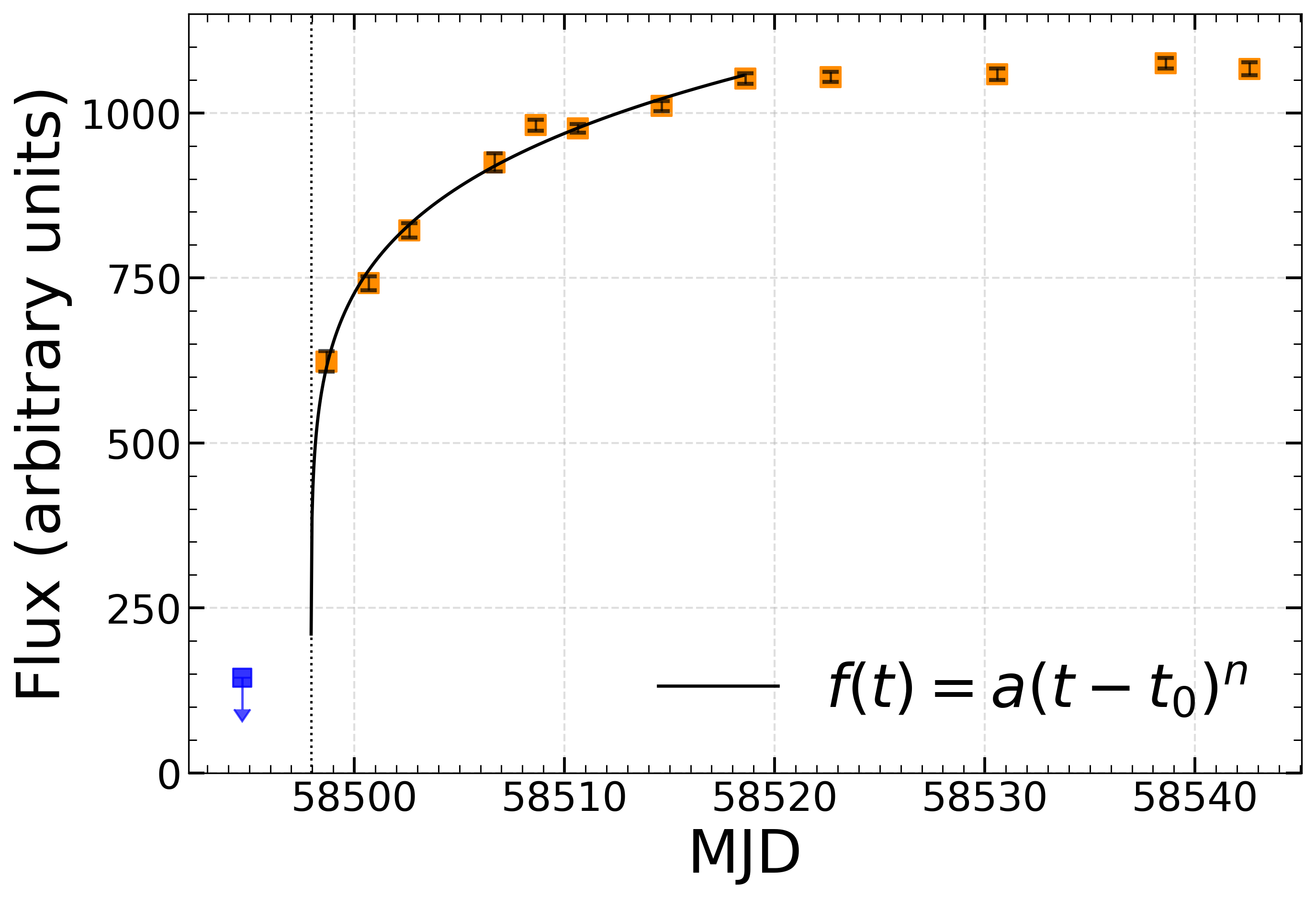}
    \caption{Estimate of the explosion epoch. Data points are denoted by orange squares. A blue square with a downward arrow denotes the upper limit. A power-law formula (black line) is fitted to the data points in the rising phase (the upper limit is not included) to estimate the explosion epoch, $t_0$, which is denoted by the dashed line.}
    \label{fig:rising}
\end{figure}

As shown in Figure \ref{fig:rising}, we also try to determine the explosion epoch by fitting $f(t)=a(t-t_0)^n$ (e.g. \citealt{Gonzalez-Gaitan2015}) to the ATLAS $o$-band data taken during the rising phase of the light curve, where $f(t)$ is the flux at the epoch of $t$, while $a$ and $n$ are free parameters. With this method, we estimate the explosion epoch as MJD=58497.9$\pm$0.4. \\

These two results are consistent with each other within the quoted errors. To be more cautious, we choose MJD=58496.7$\pm$2.0 as the explosion epoch of SN 2019va in the following analysis.

\subsection{Comparison with other SNe II in the $V$ band}
\label{subsec:compLCs}
\begin{figure*}
	\includegraphics[width=2\columnwidth]{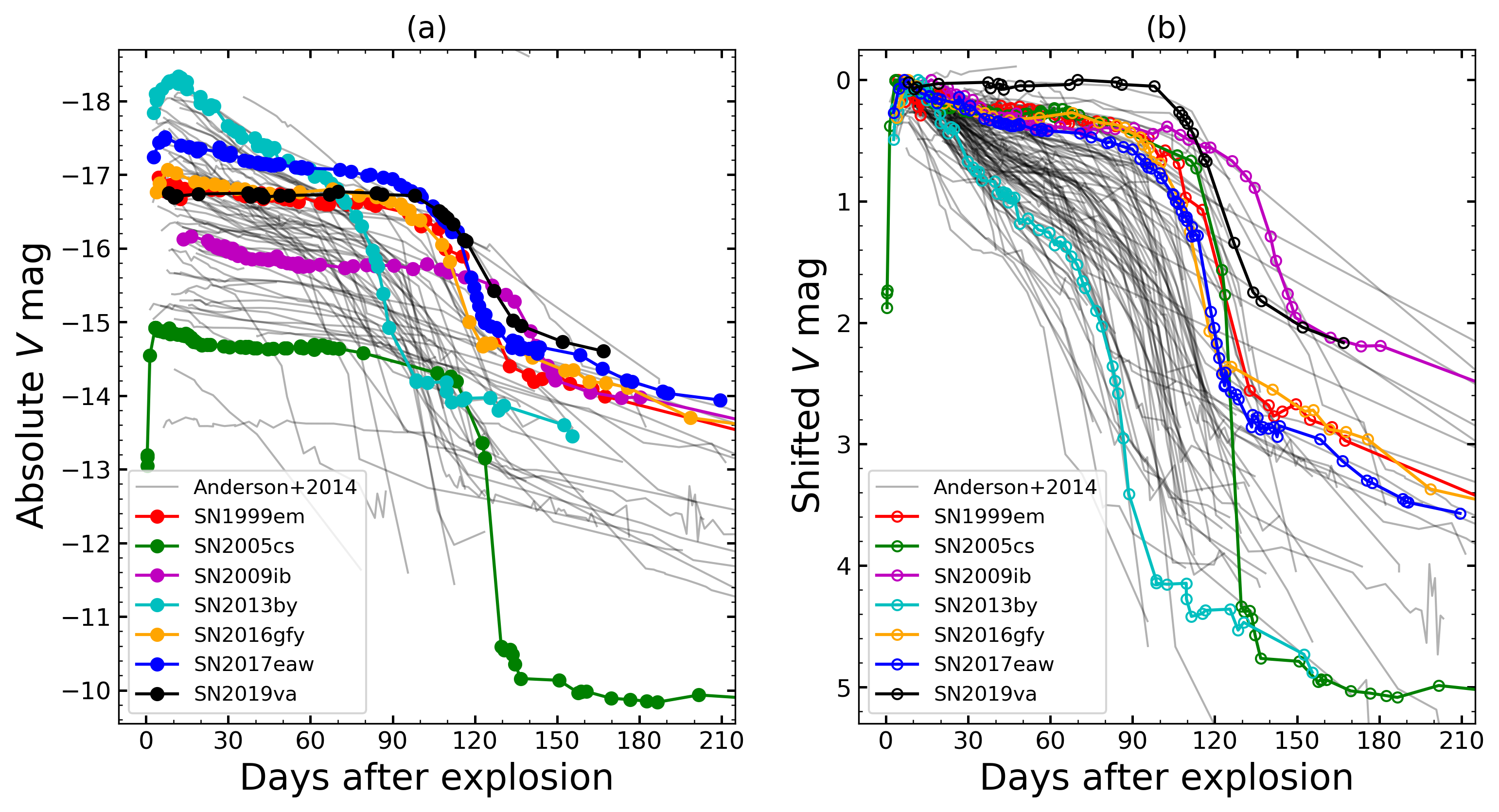}
    \caption{Comparison between the $V$-band light curves of SN 2019va and those of SN 1999em, SN 2005cs, SN 2009ib, SN 2016gfy, SN 2017eaw, and the SN-II sample of \citet{Anderson2014}. \textit{panel (a)}: extinction-corrected absolute $V$-band light curves. \textit{panel (b)}: $V$-band light curves shifted vertically to their maximum.}
    \label{fig:compVband}
\end{figure*}

In Figure \ref{fig:compVband}, we show the $V$-band light curve of SN 2019va and those of the comparison sample. Within these SNe II, SN 1999em is the prototype SN IIP. The light curves of SN 2005cs and SN 2009ib both show a very flat plateau feature, which is also seen in SN 2019va. However, the plateau luminosity of these three SNe II are very different. SN 2013by is presented here to show readers that SN 2019va does not belong to the class of SNe II with large post-peak decline rates, which historically were classified as SNe IIL. SN 2017eaw is included since it is a well studied SN II with a large amount of data. SN 2016gfy is included because it shows many similarities to SN 2019va. We also include the $V$-band light curves of the SN-II sample from \citet{Anderson2014} to demonstrate that SNe IIP and SNe IIL form a continuous distribution.\\ 

\begin{figure*}
	\includegraphics[width=2\columnwidth]{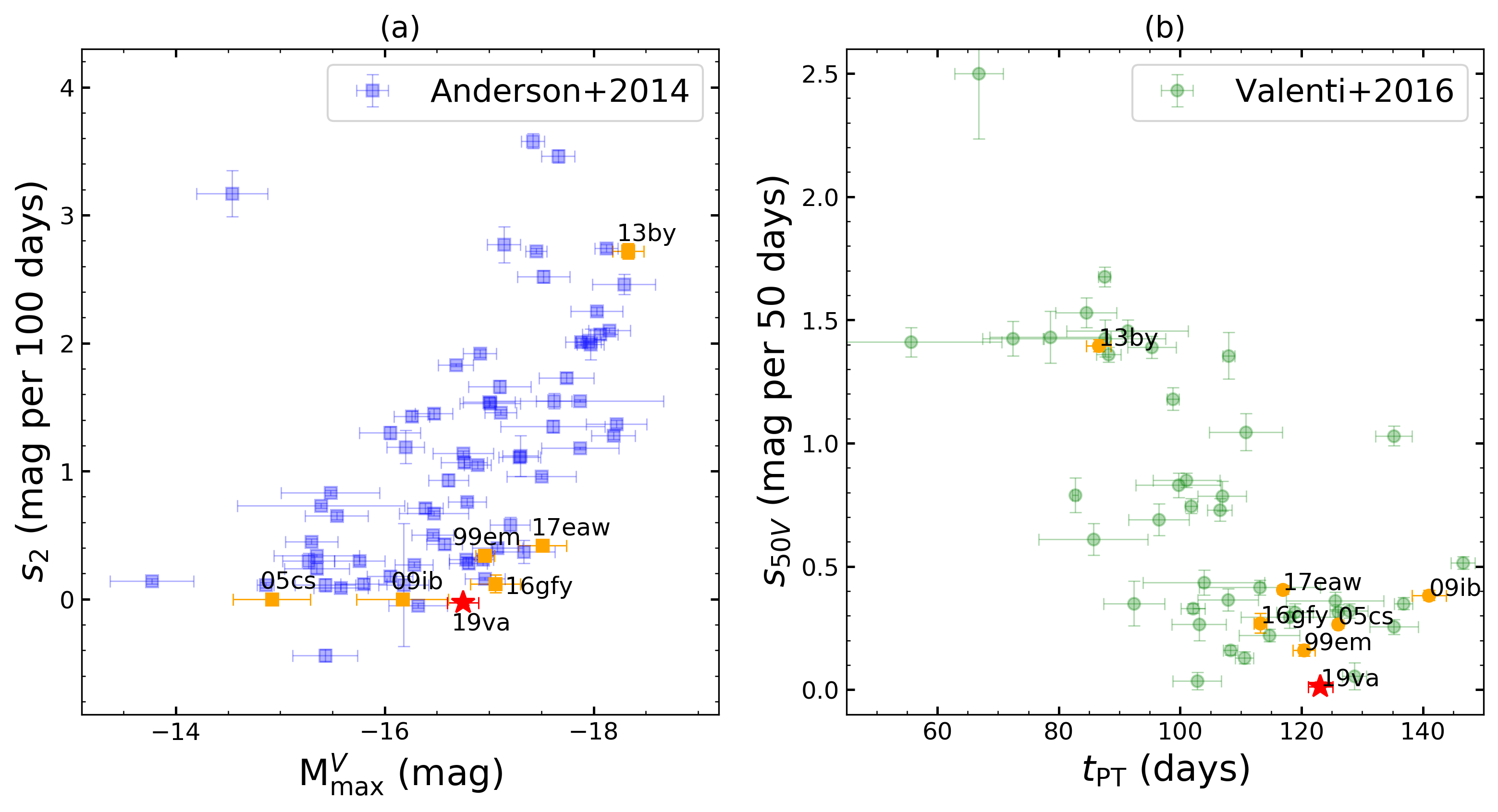}
    \caption{Parameters measured from the $V$-band light curves of SNe II. \textit{panel (a)}: correlation between the plateau decline rate ($s_2$) and the absolute $V$-band peak magnitude ($M_{\rm max}^{V}$) proposed by \citet{Anderson2014}. Blue squares represent the sample collected from \citet{Anderson2014}. The comparison SNe are denoted by orange squares and SN 2019va is denoted by a red star. \textit{panel (b)}: magnitude decline rate measured within 50 days after $V$-band peak ($S_{50V}$), versus duration between the explosion epoch and the midpoint of the transition phase ($t_{\rm PT}$). The sample collected by \citet{Valenti2016} are shown by green circles. The comparison SNe II are represented by orange circles, while SN 2019va is denoted with a red star.}
    \label{fig:compVpar}
\end{figure*}

\citet{Anderson2014} defined a quantity, $s_2$ [in units of mag\,(100 days)$^{-1}$], as the plateau slope between $t\approx30$d and $t\approx100$d after explosion to describe the decline rate of the light curve during the plateau phase. Note that the magnitude decline rate measured before $t\approx30$d is dubbed as $s_1$ by \citet{Anderson2014}. Usually, $s_1$ is larger than $s_2$. For SN 2019va, however, we do not see such a transition from $s_1$ to $s_2$. By fitting a straight line to the $V$-band data between $t=0$\,d and $t=100$\,d, we yield $s_2=-0.02\pm0.02$\,mag\,(100\,d)$^{-1}$ for SN 2019va. This $s_2$ value indicates that the plateau of SN 2019va is rather flat, as shown in Figure \ref{fig:compVband}(b). The first $V$-band data point of SN 2019va was taken at $t=8.2$\,d after explosion with an apparent magnitude of 16.50$\pm$0.02, while the brightest value in the $V$ band is measured as 16.48$\pm$0.02\,mag at $t=70$\,d. These two values are not distinguishable within the errors. Therefore, for convenience, we take $t=8.2$\,d as the peak epoch of SN 2019va in the $V$ band. Figure \ref{fig:compVpar}(a) displays the correlation between $s_2$ and the absolute $V$-band peak magnitude proposed by \citet{Anderson2014}, where SNe II with more luminous $V$-band peaks tend to have larger decline rates during the plateau phase. However, this correlation shows large scatter. For instance, SN 2019va has a very small $s_2$ value similar to SN 2005cs and SN 2009ib, while these three SNe II have very different peak luminosities.\\

Figure \ref{fig:compVpar}(b) shows the correlation between the plateau slope and the plateau length, with a longer plateau corresponding to a larger plateau slope, which was first proposed by \citet{Valenti2016}. To describe the plateau slope, they defined a new quantity, $S_{50V}$, which is calculated by fitting a straight line to the $V$-band data spanning from the peak epoch to 50 days after that. To assess the plateau length, they measured $t_{\rm PT}$ by fitting the transition-phase $V$-band light curves of SNe II with the function
\begin{equation}
    \label{eq:tranfit}
    m(t) = \frac{-a_0}{1+e^{(t-t_{\rm PT})/\omega_0}} + p_0\frac{t-t_{\rm PT}}{100\rm d} + m_0.
\end{equation}
The above function was first proposed by \citet{Olivares2010}, where $t_{\rm PT}$ is the midpoint of the transition phase, $a_0$ represents the dropping depth of the transition phase,  $\omega_0$ is a time scale reflecting the duration of the transition phase, and $p_0$ [in units of mag\,(100\,d)$^{-1}$] describes the decay rate of the tail-phase light curve. We measure the values of $S_{50V}$ and $t_{\rm PT}$ for SN 2019va and the comparison SNe II, and show them in Figure \ref{fig:compVpar}(b). SN 2019va has a very small $S_{50V}$ and a relatively large $t_{\rm PT}$, roughly conforming to the negative correlation found by \citet{Valenti2016}. 

\subsection{Colour evolution}
\label{subsec:color}

\begin{figure}
	\includegraphics[width=\columnwidth]{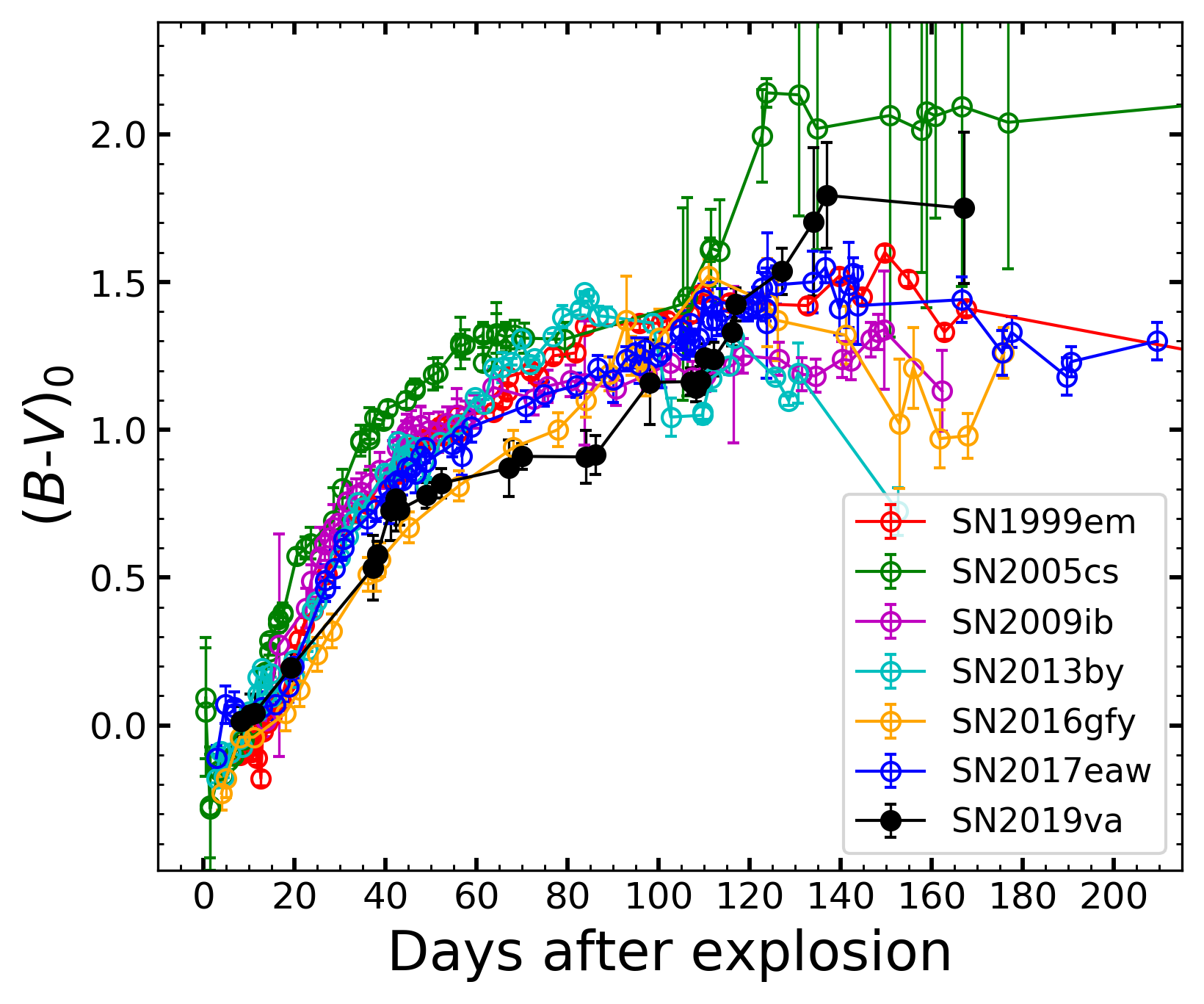}
    \caption{Intrinsic $B-V$ colour evolution of SN 2019va compared with those of some representative SNe II.}
    \label{fig:compBVcolor}
\end{figure}

We compare the intrinsic $B-V$ colour evolution of SN 2019va with those of the comparison SNe II in Figure \ref{fig:compBVcolor}. The colours of these SNe II become progressively red during the plateau phase due to the continuous decrease in the photospheric temperature after explosion. However, SN 2016gfy and SN 2019va are bluer than the other SNe II during the phase from $t\approx 30$d to $t\approx 100$d. \citet{Singh2019} also noticed this bluer evolution in their study of SN 2016gfy, and they attributed it to the weakened line blanketing effect in the blue region of the spectrum. They inferred a sub-solar metallicity ($\sim0.6Z_{\odot}$) for the environment of SN 2016gfy. From the analysis presented in Section \ref{sec:host}, a low metallicity of $\lesssim0.5Z_{\odot}$ is inferred for the progenitor of SN 2019va. Given these similarities, and the fact that no prominent signatures of other energy sources (such as circumstellar interaction) are found, the blue colour of SN 2019va may be also related to the low metallicity of the progenitor star.\\

\begin{figure}
	\includegraphics[width=\columnwidth]{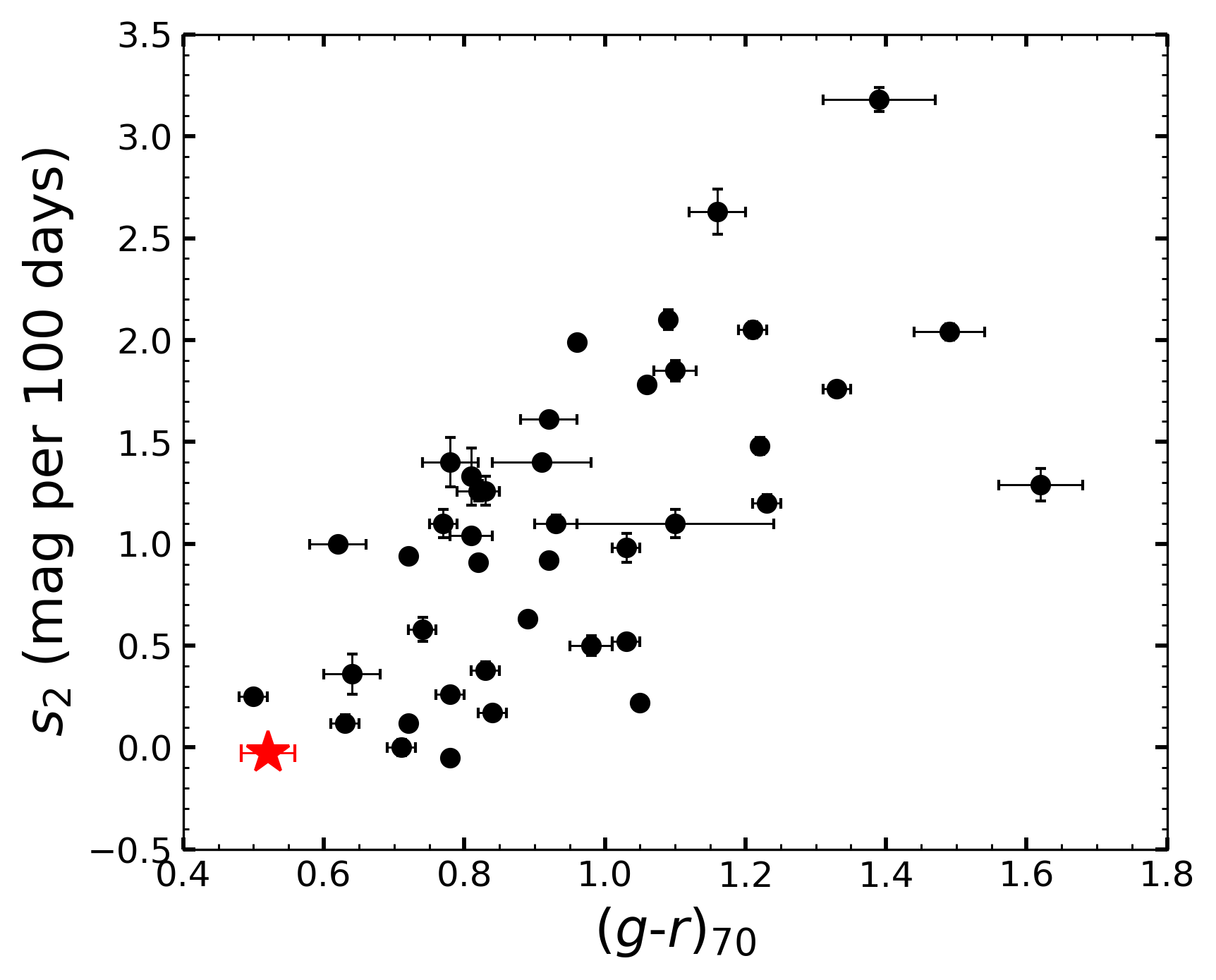}
    \caption{Correlation between the plateau slope ($s_2$) and the intrinsic $g-r$ colour at $t=70$\,d. The black dots represent the sample from \citet{deJaeger2018}, while the red star denotes SN 2019va. }
    \label{fig:gr70_s2}
\end{figure}

In Figure \ref{fig:gr70_s2}, we further compare the $g-r$ colour at $t=70$\,d and the $s_2$ parameter between SN 2019va and the SN-II sample from \citet{deJaeger2018}. As it can be seen, SN 2019va has a bluer $g-r$ colour at $t=70$\,d than most of the sample, which is consistent with its blue $B-V$ colour discussed above. Since SN 2019va also has a small value of $s_2$, it follows the trend that SNe II with smaller plateau slopes tend to have bluer $g-r$ colours at $t=70$\,d.

\subsection{The bolometric light curve}
\label{subsec:bolLC}

\begin{figure}
	\includegraphics[width=\columnwidth]{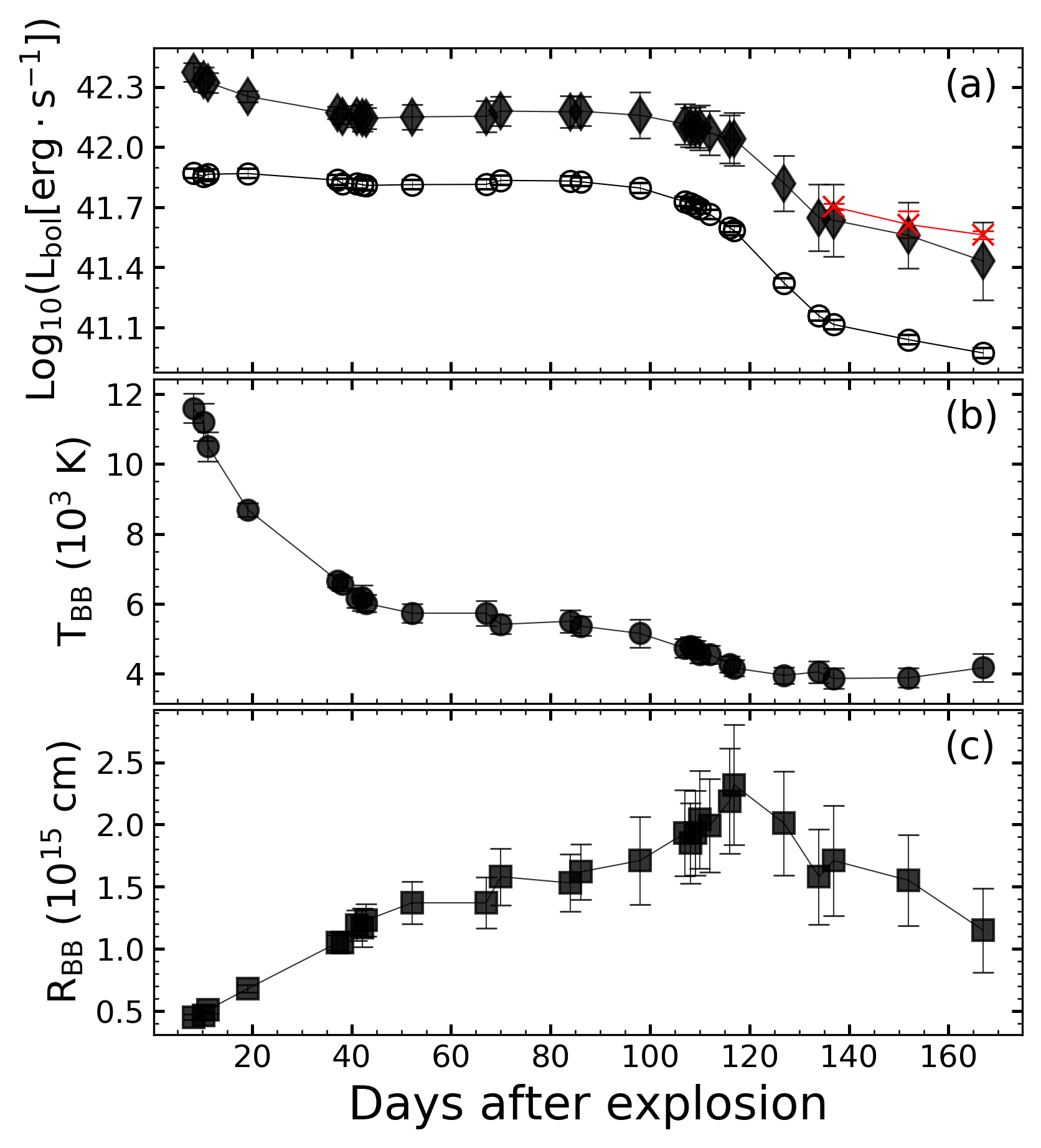}
    \caption{(\textit{a}): The bolometric light curve of SN 2019va constructed by fitting a blackbody to the SEDs (black diamonds), pseudo-bolometric light curve of SN 2019va constructed by intergrating the flux over $BVgri$-band data (unfilled circles), and tail luminosity calculated through the bolometric correction method (red ``x'' symbols). (\textit{b}): evolution of the best-fit blackbody temperature. (\textit{c}): evolution of the best-fit blackbody raidus.}
    \label{fig:bolLC}
\end{figure}

Based on the multi-band photometry, we construct the bolometric light curve of SN 2019va using a Python tool \textsc{superbol} \citep{Nicholl2018}, as shown in Figure \ref{fig:bolLC}. In \textsc{superbol}, the absolute $BVgri$-band magnitudes are first transferred to the monochromatic flux at the corresponding effective wavelength. Then two different processes are performed to establish the spectral energy distributions (SEDs) and construct the luminosity. The first is to integrate the optical flux trapezoidally to build the so-called \textbf{pseudo}-bolometric light curve; the second is to fit the optical flux with a blackbody spectrum, and then integrate the blackbody curve, so that the resultant bolometric light curve includes the additional blackbody correction. Note that in the latter process, the flux below 3000\,\AA~ is not integrated directly with the blackbody value. Instead, it is multiplied by a factor, ($\lambda$/3000\,\AA)$^{1.0}$ before integration.\footnote{refer to \url{https://github.com/mnicholl/superbol}} This is because the SN radiation is suppressed in the ultraviolet (UV) region due to the formation of the line forest from Fe-group elements when the SN ejecta expands and cools, which is known as the line blanketing effect \citep{Baron1996, Lusk2017}. This factor is used to mimic the line blanketing effect. During the tail phase of the light curves, spectra are gradually dominated by emission lines, and the continuum becomes weak and flat. As a result, a blackbody fit usually brings large uncertainties, as shown by the large error bars of the tail-phase luminosity in Figure \ref{fig:bolLC}.\\

\citet{Bersten2009} proposed that the tail-phase $V$-band magnitude can be converted to the bolometric luminosity by adding a bolometric correction term. Based on the densely-sampled, high-quality data of SN 1987A, SN 1999em and SN 2003hn, they proposed an empirical formula,
\begin{equation}
    \label{eq:bolcor}
    \log_{10}L_{\rm bol} = -0.4 \times [BC + V - A_{V} + 11.64] + \log_{10}(4\pi D^2),
\end{equation}
where $L_{\rm bol}$ is the estimated bolometric luminosity in units of erg\,s$^{-1}$, $BC$ is the bolometric correction term with a value of -0.7 mag, $V$, $A_{V}$, and $D$ are the $V$-band magnitude, extinction value, and distance to the supernova (in units of cm), respectively. Following this empirical formula, we estimate the tail-phase bolometric luminosity for SN 2019va, as displayed in Figure \ref{fig:bolLC} by red ``x'' symbols.\\

\begin{figure}
	\includegraphics[width=\columnwidth]{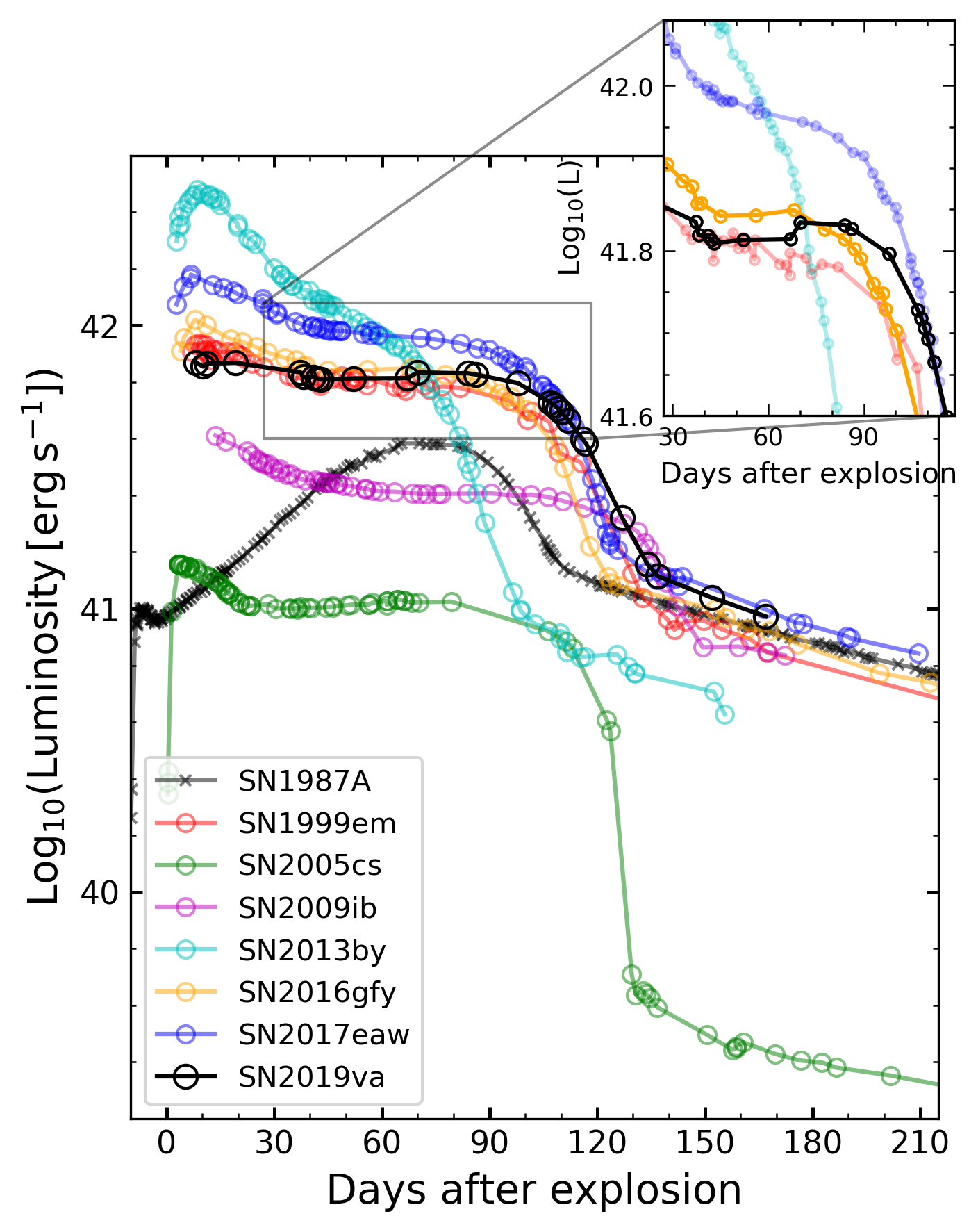}
    \caption{Pseudo-bolometric light curves of SN 2019va and some well-observed SNe II. Details of some SNe II are zoomed in to show the weak brightening in the late plateau of SN 2016gfy and SN 2019va.}
    \label{fig:compBolLC}
\end{figure}

To make consistent comparisons, we collect $BVRI$- or $BVgri$-band magnitudes of some SNe II from the literature (including SN 1987A, SN1999em, SN 2005cs, SN2009ib, SN 2013by, SN 2016gfy, and SN 2017eaw), and generate pseudo-bolometric\footnote{i.e., we directly integrate the monochromatic flux transferred from the optical-band photometric data for each epoch, without applying any blackbody corrections.} light curves for these SNe II using \textsc{superbol}. In Figure \ref{fig:compBolLC}, we compare the pseudo-bolometric light curves of SN 2019va and the above SNe II. We find that SN 2016gfy and SN 2019va both show a possible weak brightening in the late plateau phase, which is also seen in SNe II with lower plateau luminosity, namely SN 2005cs and SN 2009ib, but not obviously seen in SNe II with comparable or brighter plateau luminosity, such as SN 1999em and SN 2017eaw. This phenomenon was also noticed by \citet{Singh2019}, and they dubbed it as a ``bump'' in the late plateau phase ($t=50\sim100$\,d) of SN 2016gfy. The reason for this weak brightening will be discussed in Section \ref{subsec:late-plateau}. During the tail phase, we find SN 2019va is slightly brighter than SN 1987A, indicating a possible larger mass of $^{56}$Ni synthesized in the explosion of SN 2019va.\\

\section{Spectroscopic analysis}
\label{sec:spec}

\subsection{Comparison with other SNe II}
\label{subsec:compspec}
\begin{figure*}
	\includegraphics[width=2\columnwidth]{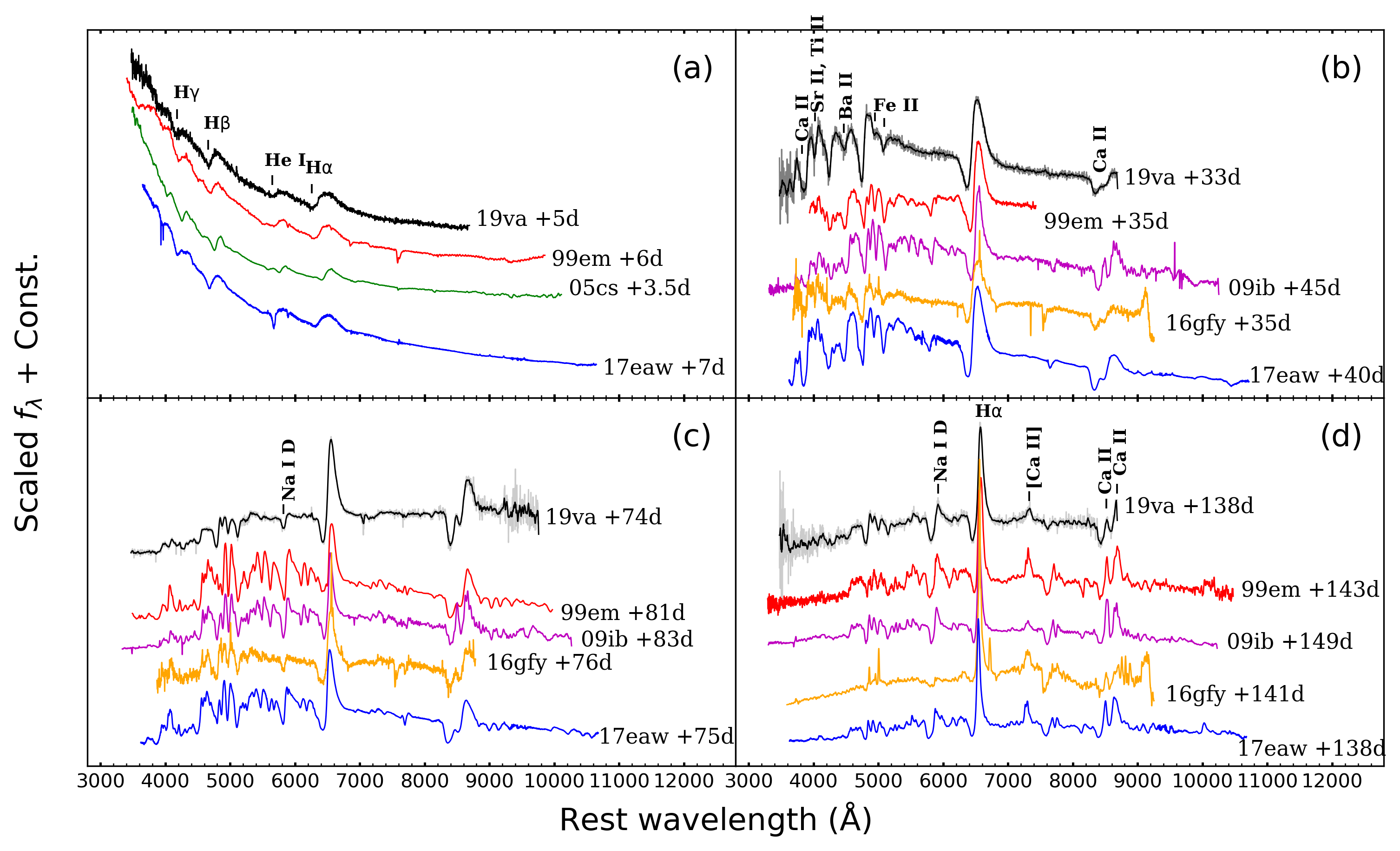}
    \caption{Comparison of spectral evolution between SN 2019va and some well-studied SNe II at similar epochs. All spectra are corrected for the reddening and the host-galaxy redshift. The epochs marked besides the spectra are relative to the explosion date. Important spectral lines are labelled.}
    \label{fig:compspec}
\end{figure*}
In Figure \ref{fig:compspec}, we compare the spectra of SN 2019va with those of SN 1999em, SN 2005cs, SN 2009ib, SN 2016gfy, and SN 2017eaw at four different phases. In the early phase (i.e., $t<10$\,d; see panel \textit{a}), the spectra all show very blue continua, along with broad P Cygni profiles of the Balmer series. For SN 2019va, we notice that at the wavelength of $\sim$5700\,\AA, a broad P Cygni profile exists in the $t=5$\,d spectrum, but it disappears in the later spectra until an absorption trough emerges at this location again in the $t\approx60\,$d spectrum. The former P Cygni profile is identified as \ion{He}{i} $\lambda5876$ considering the high temperature, while the latter absorption trough can be attributed to \ion{Na}{i~D}. At $t\approx1\,\rm month$, when the SNe enter the early plateau phase (see panel \textit{b} of Figure \ref{fig:compspec}), metal absorption lines emerge in their spectra, as a result of the decrease in the photospheric temperature. Through comparison with other SNe II, we identify the presence of \ion{Ca}{ii}, \ion{Fe}{ii}, \ion{Ba}{ii} lines, and possibly blended features of \ion{Sr}{ii} and \ion{Ti}{ii} for SN 2019va. During this phase, the \ion{Na}{i} absorption feature emerges in the spectra of SN 1999em, SN 2009ib, and SN 2017eaw, but not in SN 2016gfy or SN 2019va. During the late plateau phase (see panel \textit{c} of Figure \ref{fig:compspec}), the \ion{Na}{i} lines become visible in the spectra of SN 2016gfy and SN 2019va, but the absorption troughs are very weak compared with other SNe II. Moreover, SN 2016gfy and SN 2019va also lack prominent metal lines at the left side of the \ion{Na}{i} line (5200\,\AA $\sim$ 6200\,\AA), which is different from the other SNe II. These features are believed to be related to the low metallicity of their progenitors \citep{Dessart2014, Anderson2016}. At $t=138\,$d (see panel \textit{d} of Figure \ref{fig:compspec}), when the light curves begin to enter the tail phase, the photospheric features become weak, while emission lines, such as [\ion{Ca}{ii}] $\lambda\lambda7291,7323$ and \ion{Ca}{ii} NIR triplet, start to emerge in the spectra.

\subsection{Expansion velocity}
\begin{figure*}
	\includegraphics[width=2\columnwidth]{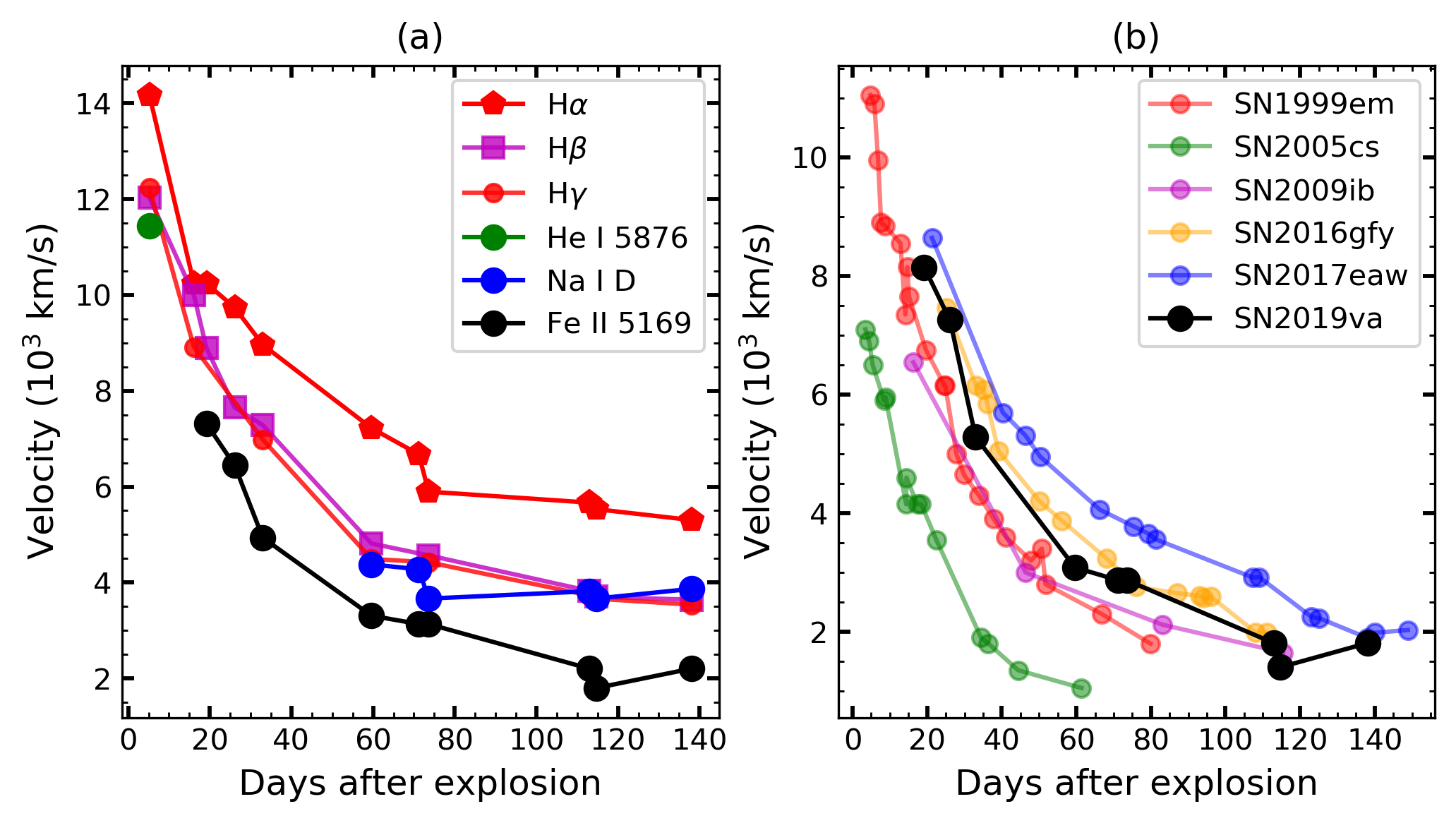}
    \caption{\textit{(a)}: velocity evolution of different species seen in the spectra of SN 2019va. \textit{(b)}: photospheric velocity of SN 2019va (transferred from \ion{Fe}{ii} $\lambda5169$) compared with that of SN 1999em, SN 2005cs, SN 2009ib, SN 2016gfy, and SN 2017eaw.}
    \label{fig:velocity}
\end{figure*}
Figure \ref{fig:velocity}(a) shows the expansion velocities of H$\alpha$, H$\beta$, H$\gamma$, \ion{He}{i} $\lambda5876$, \ion{Na}{i~D} and \ion{Fe}{ii} $\lambda5169$, which are inferred from the absorption minimum of the corresponding elements. At early epochs, the photosphere lies at the outer layers of the ejecta, therefore we see high velocities as inferred from H$\alpha$, H$\beta$ and \ion{He}{i} lines. As the photosphere recedes, metal lines emerge in the spectra. \citet{Takats2012} 
suggest that the photospheric velocity ($v_{\rm ph}$) can be derived from the velocity of \ion{Fe}{ii} $\lambda5169$ ($v_{\rm Fe}$) using the following empirical equation:  
\begin{equation}
    \label{eq:velocity}
    v_{\rm Fe}/v_{\rm ph} = a_0 + a_1\times v_{\rm Fe} + a_2\times v_{\rm Fe}^2,
\end{equation}
where $a_0=1.641$, $a_1=-2.297\times10^{-4}$, $a_2=1.751\times10^{-8}$, and velocities are in units of km\,s$^{-1}$. From this equation, the photospheric velocities of SN 2019va are determined and displayed in Figure \ref{fig:velocity}(b), together with those of SN 1999em \citep{Takats2012}, SN 2005cs \citep{Takats2012}, SN 2009ib \citep{Takats2015}, SN 2016gfy \citep{Singh2019}, and SN 2017eaw \citep{VanDyk2019}. The photospheric velocity of SN 2019va lies between those of SN 1999em and SN 2017eaw, and is comparable to that of SN 2016gfy.

\subsection{Pseudo-equivalent width of \ion{Fe}{ii} $\lambda$5018 and metallicity}
\label{subsec:pEW_FeII}

\begin{figure}
	\includegraphics[width=\columnwidth]{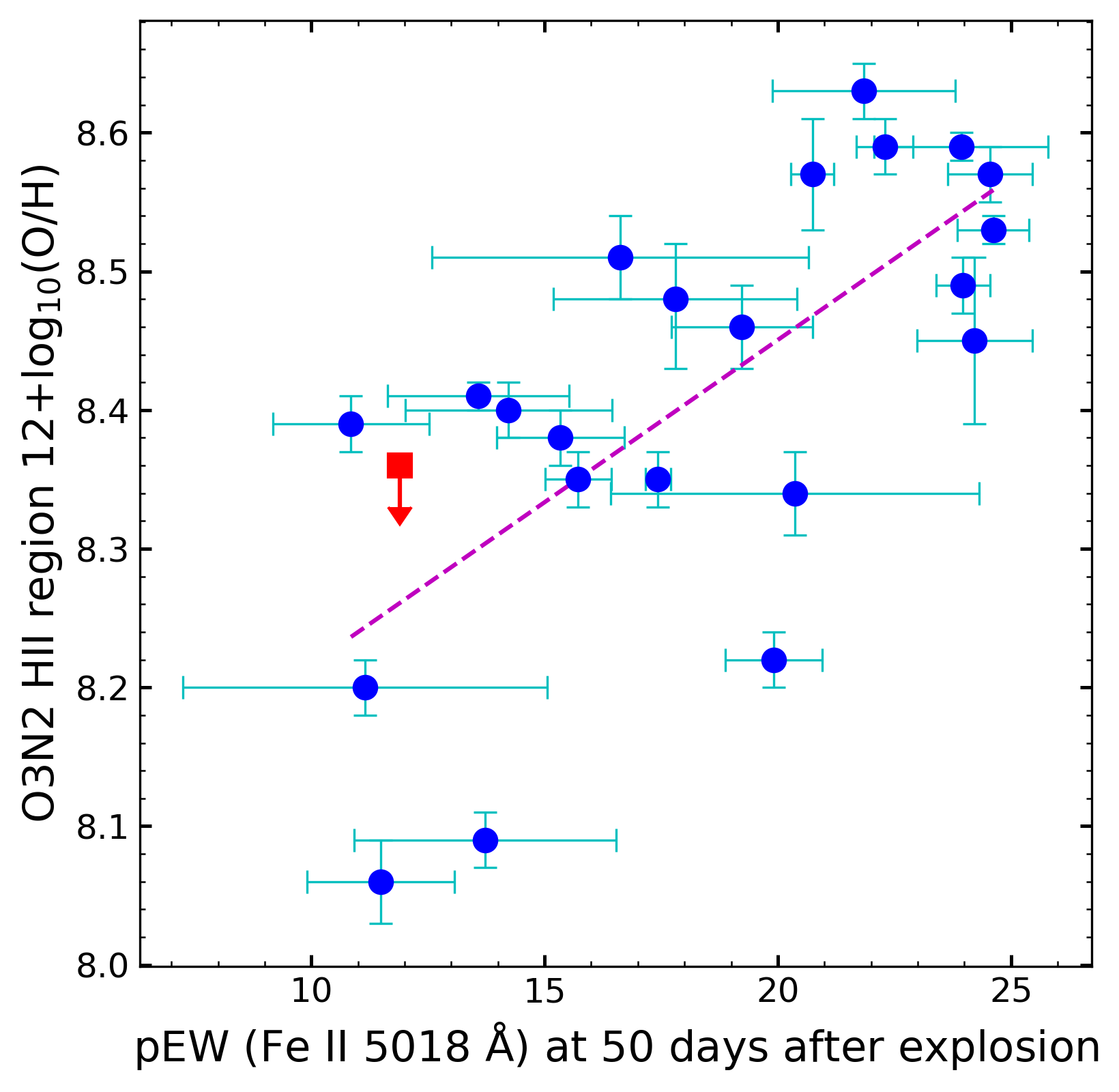}
    \caption{Correlation between pEW of \ion{Fe}{ii} $\lambda$5018 at $t=50$\,d after explosion and environment metallicity inferred from emission line flux ratio (O3N2). Blue dots represent the ``Gold'' sample from \citet{Anderson2016}, while the downward arrow represents the upper limit derived for SN 2019va. The dashed line indicates the trend for such a correlation. }
    \label{fig:pEW}
\end{figure}

As discussed in Section \ref{subsec:compspec}, when comparing the spectra of SN 2019va to those of other SNe II, we find that SN 2019va shows very weak metal lines. Through spectral modelling, \citet{Dessart2014} found that a low metallicity progenitor would result into a small pseudo-equivalent width (pEW) of \ion{Fe}{ii} $\lambda$5018, which is confirmed in the sample of \citet{Anderson2016}. In particular, when restricting the sample to those with plateau slope ($s_2$) less than 1.5\,mag\,(100d)$^{-1}$ and optically thick phase duration (OPTd) larger than 70 days (this sub-sample were dubbed as the ``Gold'' sample), \citet{Anderson2016} found that a statistically significant positive correlation exists between the pEW of \ion{Fe}{ii} $\lambda$5018 at $t=50$d and the oxygen abundance of the host \ion{H}{ii} region, as shown in Figure \ref{fig:pEW}. For SN 2019va, we measure the pEW of \ion{Fe}{ii} $\lambda$5018 as 11.90$\pm$0.22\,\AA~at $t=50\rm d$. Such a weak \ion{Fe}{II} absorption is compatible with the fact that SN 2019va has a metal-poor progenitor.

\section{discussion}
\label{sec:discuss}

\subsection{The $^{56}$Ni mass}
\label{subsec:Nimass}
An important parameter, which is believed to control the morphology of the tail-phase light curve of SNe II, is the mass of $^{56}$Ni synthesized in the explosion. The $^{56}$Ni mass can be estimated from the bolometric luminosity during the tail phase through the following equation \citep{Hamuy2003, Zhangjujia2020},
\begin{equation}
   \label{eq:Hamuy03}
   M_{\rm Ni} = 7.866 \times 10^{-44} \times L_t \times \exp \left[ \frac{t/(1+z)-6.1}{112.26} \right],
\end{equation}
where $M_{\rm Ni}$ is the mass of $^{56}$Ni in units of solar masses, $t$ is days after explosion, $z$ is the redshift of the SN, and $L_t$ is the bolometric luminosity at time $t$ in units of erg\,s$^{-1}$. In Section \ref{subsec:bolLC}, we obtained the tail-phase bolometric luminosity converted from the $V$-band data using the ``bolometric correction'' method (see Equation (\ref{eq:bolcor}) and Figure \ref{fig:bolLC}). Substituting the luminosity values into equation (\ref{eq:Hamuy03}) and averaging the results, we obtain a $^{56}$Ni mass of 0.12$\pm$0.02\,$M_{\odot}$.\\

The $^{56}$Ni mass can be also estimated by comparing the tail-phase bolometric light curve between SN 2019va and SN 1987A. In Section \ref{subsec:bolLC} (also see Figure \ref{fig:compBolLC}), we have constructed the pseudo-bolometric light curves of these two SNe II by integrating the monochromatic flux over optical bands (i.e., $BVRI$ or $BVgri$ bands). We find the luminosity of SN 2019va is 1.17$\pm$0.04 times larger than that of SN 1987A at the radioactive tail ($t>135$\,d). Since the $^{56}$Ni mass for SN 1987A is estimated as 0.075$\pm$0.015\,$M_{\odot}$ by \citet{Arnett_Fu1989}, we then estimate the $^{56}$Ni mass of SN 2019va as 0.088$\pm$0.018\,$M_{\odot}$. When we compare the pseudo-bolometric light curves of SNe II in Figure \ref{fig:compBolLC}, we find the tail luminosity of SN 2019va is similar to that of SN 2017eaw, therefore they are expected to have a similar $^{56}$Ni mass. Indeed, SN 2017eaw is estimated to produce a $^{56}$Ni mass of $\sim 0.08$\,$M_{\odot}$ in the SN explosion \citep{Rho2018, VanDyk2019}.\\  

The first method used to estimate the $^{56}$Ni mass is based solely upon $V$-band data, while the second one relys on all optical bands ($BVgri$ for SN 2019va). In Figure \ref{fig:compVband}(a), SN 2019va has a slightly more luminous tail-phase light curve than SN 2017eaw in the $V$ band. However, their pseudo-bolometric light curves show comparable tail luminosity, as seen in Figure \ref{fig:compBolLC}. This is possibly because the tail-phase light curves of SN 2019va in the other optical bands are not as luminous as in the $V$ band. For example, SN 2019va shows a much redder $B-V$ colour in the tail phase ($t>135$\,d) relative to most comparison SNe II (see Figure \ref{fig:compBVcolor}), illustrating that SN 2019va is unusually faint in $B$ band. Such photometric behaviour of SN 2019va during the tail phase means that the true bolometric colour correction of SN 2019va is possibly lower than the one used in the first method, which may be the reason that the $^{56}$Ni mass estimated from the first method appears larger than the second one. \\

\begin{figure}
	\includegraphics[width=\columnwidth]{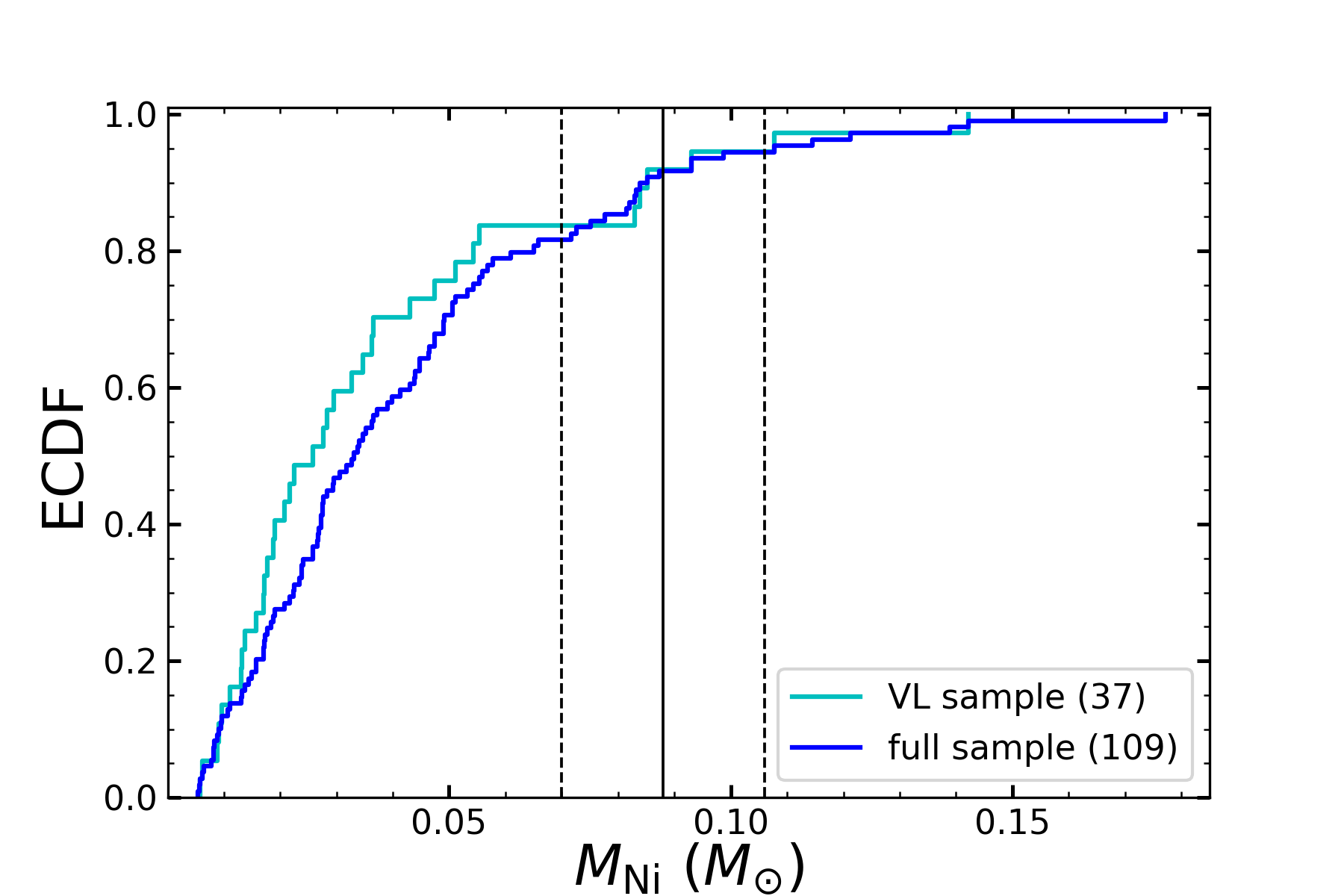}
    \caption{The empirical cumulative distribution function (ECDF) for the $^{56}$Ni mass of SNe II obtained by \citet{Rodriguez2021}. ``VL sample'' in the caption represents the volume-limited sample at distance modulus $\leq 31.2$\,mag, while the ``full sample'' represents the full sample without a distance cut. The numbers in the parentheses denote the sample number. The $^{56}$Ni mass of SN 2019va is marked by the solid line. The dashed lines denote $1\sigma$ error.}
    \label{fig:ECDF}
\end{figure}

\begin{figure}
	\includegraphics[width=\columnwidth]{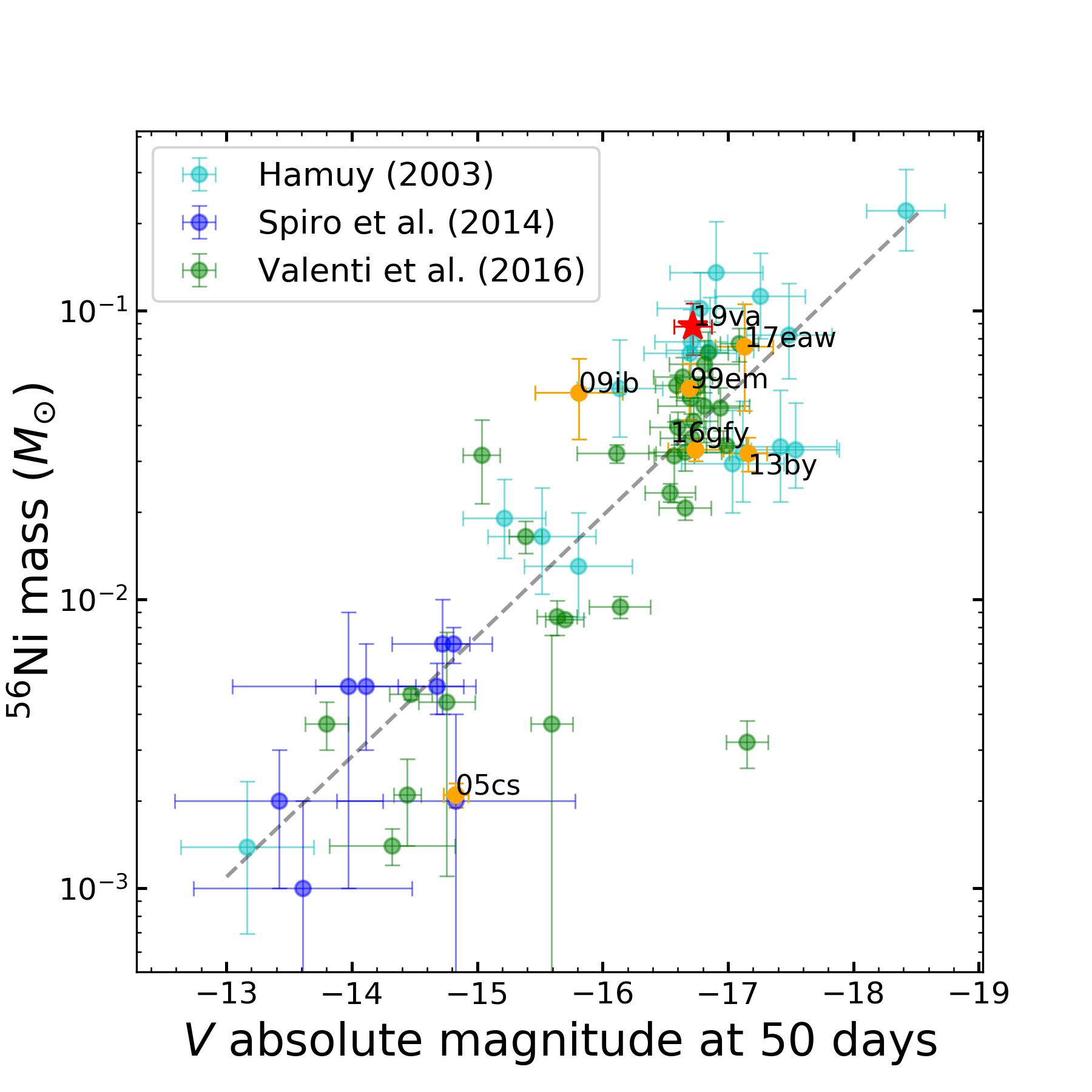}
    \caption{Correlation between $^{56}$Ni mass (in logarithm) and the absolute $V$-band magnitude at $t=50$\,d found by \citet{Hamuy2003}, \citet{Spiro2014}, and \citet{Valenti2016}. Comparison SNe II used in this paper are denoted by orange circles, while SN 2019va is marked by a red star. The dashed line is shown to guide the correlation.}
    \label{fig:M50V_Mni}
\end{figure}

In the following discussion, we adopt the $^{56}$Ni mass estimated from the second method, i.e., 0.088$\pm$0.018\,$M_{\odot}$, as it is based on multi-band data. It should be pointed out that this $^{56}$Ni mass is a relatively large value for a SN II. As a comparison, a selection-bias-corrected average of the $^{56}$Ni mass is estimated as 0.037$\pm$0.005\,$M_{\odot}$ for SNe II \citep{Rodriguez2021}. As shown in Figure \ref{fig:ECDF}, the $^{56}$Ni mass synthesized in the explosion of SN 2019va exceeds $\sim$90\% of the sample included in \citet{Rodriguez2021}. Figure \ref{fig:M50V_Mni} shows that more luminous SNe II tend to produce more $^{56}$Ni \citep{Hamuy2003, Spiro2014, Valenti2016}, and SN 2019va seems to follow this tendency, though this correlation has relatively large scatter.

\subsection{Influence of $^{56}$Ni decay on the plateau-phase light curve}
\label{subsec:late-plateau}
The pseudo-bolometric light curve of SN 2019va shows a weak brightening at the end of the plateau phase, which is similar to that of SN 2016gfy. In studying SN 2016gfy, \citet{Singh2019} attributed the weak brightening to the contribution of $^{56}$Ni decay (see their Section 8.2). Therefore, we investigate the possibility that $^{56}$Ni decay affects the plateau-phase light curve of SN 2019va.\\

It is generally believed that $^{56}$Ni decay determines the tail-phase light curve of SNe II, while the plateau-phase evolution is dominated by the emission of envelope cooling  (e.g. \citealt{Litvinova1985, Nagy2014}). However, \citet{Nakar2016} and \citet{Kozyreva2019} emphasized that $^{56}$Ni decay may play an important role in shaping the plateau-phase light curves of SNe II. They found that the energy contributed by $^{56}$Ni decay can extend the plateau duration or flatten the plateau evolution, depending on the degree of $^{56}$Ni mixing. If most $^{56}$Ni is concentrated in the centre of the ejecta, radioactive photons will diffuse out after the recombination of H/He to extend the plateau duration. In this case, the extension is not necessary smooth, and one may see a bump feature emerging in the late plateau phase (e.g., see Figures 2 and 3 in \citealt{Kozyreva2019}). However, it is possible that Rayleigh-Taylor instability or asymmetry in the exploding core brings some $^{56}$Ni into the envelope \citep{Kifonidis2003, Kasen2009}. In this case, photons contributed by $^{56}$Ni decay will be released and added to the emission from the cooling envelope at an earlier phase, which will flatten the light curve. In practice, $^{56}$Ni decay has influences on both the slope and the duration of the light-curve plateau.\\

\citet{Nakar2016} proposed an observable to quantitatively assess the influence of $^{56}$Ni decay on the light curve during the plateau phase, i.e.,
\begin{equation}
    \label{eq:etaNi}
    \eta_{\rm Ni} = \frac{\int_0^{t_{\rm Ni}} t Q_{\rm Ni}(t) \dif t}{\int_0^{t_{\rm Ni}} t (L_{\rm bol}(t) - Q_{\rm Ni}(t)) \dif t},
\end{equation}
where $L_{\rm bol}(t)$ is the bolometric luminosity at time $t$, $t_{\rm Ni}$ is the epoch separating the photospheric phase and the tail phase, and $Q_{\rm Ni}(t)$ is the instantaneous energy released by $^{56}$Ni decay, which is expressed as follows,
\begin{equation}
    \label{eq:Nidecay}
    Q_{\rm Ni}(t) = \frac{M_{\rm Ni}}{ M_{\odot}} (6.45\,\rm e^{\frac{-\it t}{8.8\,\rm days}} + 1.45\,\rm e^{\frac{-\it t}{111.3\,\rm days}}) \times 10^{43} \,\rm erg\,s^{-1},
\end{equation}
where $M_{\rm Ni}$ is the $^{56}$Ni mass. Equation (\ref{eq:etaNi}) can be interpreted as the ratio between the energy contribution from $^{56}$Ni decay and that from envelope cooling. Therefore, a larger $\eta_{\rm Ni}$ value means a larger influence of $^{56}$Ni decay on the plateau-phase light curve.\\

Using the above method, \citet{Nakar2016} obtained $\eta_{\rm Ni}$ values for 24 SNe II. In addition to this sample, we further collect 14 SNe II from the literature, for which the $\eta_{\rm Ni}$ value can be calculated. For these 14 SNe II and SN 2019va, we first construct their pseudo-bolometric light curves by integrating the monochromatic flux transferred from the $BVRI$- or $BVgri$-band data for each epoch using \textsc{superbol}. Then the epoch when SNe enter the tail phase is estimated as $t_{\rm Ni}$. We then fit the data within $\sim$10\,days after $t_{\rm Ni}$ with Equation (\ref{eq:Nidecay}) to determine $Q_{\rm Ni}(t)$ \footnote{a best-fit $M_{\rm Ni}$ will be obtained, but note that it does not represent the true $^{56}$Ni mass, because the pseudo-bolometric light curves are used.}. Finally, we calculate the $\eta_{\rm Ni}$ value for these 15 SNe II following Equation (\ref{eq:etaNi}). The results, together with the measurements from \citet{Nakar2016}, are listed in Table \ref{tab:etaNi}.\\

\begin{table}
	\centering
	\caption{The $\eta_{\rm Ni}$ parameter estimated in this paper and in \citet{Nakar2016}.}
	\label{tab:etaNi}
	\begin{threeparttable}
	\begin{tabular}{ccccc} 
		\hline
		\hline
		SNe II   &  $t_{\rm Ni}$ & ``$M_{\rm Ni}$''$^{*}$  & $\eta_{\rm Ni}$  & L type$^{**}$\\
		         &  (days)       & $M_{\odot}$             &                  &       \\
		\hline
		\hline
		\multicolumn{5}{c}{\textit{a}. Results calculated in this paper.}\\
		\hline
        SN 2001X  &  125        & 0.0227  & 0.65  & $BVRI$ \\
        SN 2009kr &  102        & 0.0098  & 0.62  & $BVRI$ \\
        SN 2013bu &  113        & 0.0011  & 0.12  & $BVgri$ \\
        SN 2013fs &  105        & 0.0192  & 0.47  & $BVRI$  \\
        LSQ13dpa  &  145        & 0.0332  & 0.73  & $BVgri$ \\
        SN 2014cy &  130        & 0.0025  & 0.13  & $BVgri$ \\
        SN 2014dw &  98         & 0.0053  & 0.19  & $BVgri$ \\
        SN 2014G  &  97         & 0.0209  & 0.38  & $BVRI$  \\
        ASASSN14dq  & 114       & 0.0223  & 0.44  & $BVgri$ \\
        ASASSN14gm  & 120       & 0.0332  & 0.64  & $BVgri$ \\
        ASASSN14ha  & 144       & 0.0006  & 0.08  & $BVgri$ \\
        SN 2015W    & 127       & 0.0178  & 0.54  & $BVgri$ \\
        SN 2016gfy  & 126       & 0.0259  & 0.60  & $BVRI$ \\
        SN 2017eaw  & 127       & 0.0317  & 0.50  & $BVRI$ \\
        SN 2019va   & 137       & 0.0323  & 0.80  & $BVgri$ \\
		\hline
		\hline
		\multicolumn{5}{c}{\textit{b}. Results listed in Table 2 of \citet{Nakar2016}; ``optical sample''. } \\
		\hline
		SN 1992H    & 142       & 0.042   & 0.71  & $BVR$ \\
		SN 1995ad   & 98        & 0.015   & 0.4   & $UBVRI$ \\
		SN 2001dc   & 120       & 0.0022  & 0.43  & $BVRI$ \\
		SN 2003Z    & 130       & 0.0021  & 0.27  & $BVRI$\\
		SN 2004A    & 120       & 0.0175  & 0.63  & $BVRI$\\
		SN 2008in   & 115       & 0.0055  & 0.39  & $BVRI$\\
		SN 2009bw   & 140       & 0.012   & 0.17  & $UBVRI$\\
		SN 2009dd   & 130       & 0.021   & 0.46  & $UBVRI$\\
		SN 2010aj   & 94        & 0.005   & 0.09  & $UBVRI$\\
		SN 2013ab   & 116       & 0.026   & 0.47  & $uvo$ \\
		SN 2013ej   & 112       & 0.011   & 0.16  & $uvo$ \\
		\hline
		\hline
		\multicolumn{5}{c}{\textit{c}. Results listed in Table 1 of \citet{Nakar2016}; ``bolometric sample''. }\\
		\hline
        SN 1999em   & 136       & 0.047   & 0.54  & bol \\
                    & [136]$^{***}$ & [0.0219] & [0.53] & [$BVRI$] \\
        SN 1999gi   & 130       & 0.058   & 0.63  & bol \\
        SN 2003hn   & 110       & 0.032   & 0.31  & bol \\
        SN 2004et   & 136       & 0.050   & 0.43  & $UBVRIJHK$ \\
                    & (136)     & (0.025) & (0.32) & ($UBVRI$) \\
        SN 2005cs   & 130       & 0.050   & 0.26  & bol \\
                    & (130)     & (0.017) & (0.15) & ($UBVRI$) \\
        SN 2007od   & 112       & 0.020   & 0.12  & uvoir \\
        SN 2009N    & 112       & 0.020   & 0.61  & bol \\
                    & (112)     & (0.008) & (0.45) & ($BVRI$) \\
        SN 2009ib   & 142       & 0.046   & 2.6   & uvoir \\
                    & [142]     & [0.0220] & [1.8] & [$BVgri$] \\
        SN 2009md   & 121       & 0.005   & 0.28  & $UBVRIJHK$ \\
        SN 2012A    & 117       & 0.009   & 0.29  & uvoir \\
        SN 2012aw   & 135       & 0.049   & 0.49  & uvoir \\
        SN 2012ec   & 112       & 0.035   & 0.57  & $UBVRIJHK$ \\
        SN 2013by   & 104       & 0.029   & 0.2   & bol \\
                    & [104]     & [0.015] & [0.2] & [$BVgri$] \\
        \hline
		\hline
	\end{tabular}
	    \begin{tablenotes}
	        \footnotesize
	          \item[*]Note: the $M_{\rm Ni}$ is not necessary the true $^{56}$Ni mass, due to the use of the pseudo-bolometric light curve.
	          \item[**] This column lists the ways to construct the bolometric light curves. Type ``bol'' means that a bolometric correction factor is included or a blackbody is fit to the SEDs when constructing the bolometric light curve, while all other types means direct integration of the flux in the listed wave bands. For more infomation, we refer readers to \citet{Nakar2016}.
	          \item[***] data in ``[   ]'' are calculated by us.
	    \end{tablenotes}
	\end{threeparttable}
\end{table}

Within the sample of \citet{Nakar2016} (N = 24), 13 SNe II also have infrared (IR) data, resulting in the construction of more accurate bolometric light curves compared to those with only optical data. These 13 SNe II form their ``bolometric sample'' (see their Table 1 or Table \ref{tab:etaNi}\textit{c} in this paper), while the remaining 11 SNe II without IR data form the ``optical sample'' (see their Table 2 or Table \ref{tab:etaNi}\textit{b} in this paper). They argued that $\eta_{\rm Ni}$ estimated from the pseudo-bolometric light curve (i.e., without IR data) is actually very accurate because this parameter is a ratio of luminosities so that it is insensitive to the missing bolometric correction factor (see their Section 3). They performed an examination on SN 2004et, SN 2005cs, and SN 2009N, which are in the ``bolometric sample''. For these three SNe II, they built the pseudo-bolometric light curves based solely on ($U$)$BVRI$-band data, and recalculated the $\eta_{\rm Ni}$ values. They found that the results are only smaller by a moderate factor of $\lesssim 1.4$ than those estimated from the bolometric light curves (i.e., including IR data). We examine three additional SNe II in their bolometric sample (i.e., SN 1999em, SN 2009ib, and SN 2013by), finding the results support that obtained by \citet{Nakar2016} (see data marked by ``[ ]'' in Table \ref{tab:etaNi}\textit{c}.). In addition, for SN 2016gfy, \citet{Singh2019} estimated a value of $\eta_{\rm Ni}=0.6$ based on their pseudo-bolometric light curve, which is the same as our result (see Table \ref{tab:etaNi}\textit{a}).\\

\begin{figure}
	\includegraphics[width=\columnwidth]{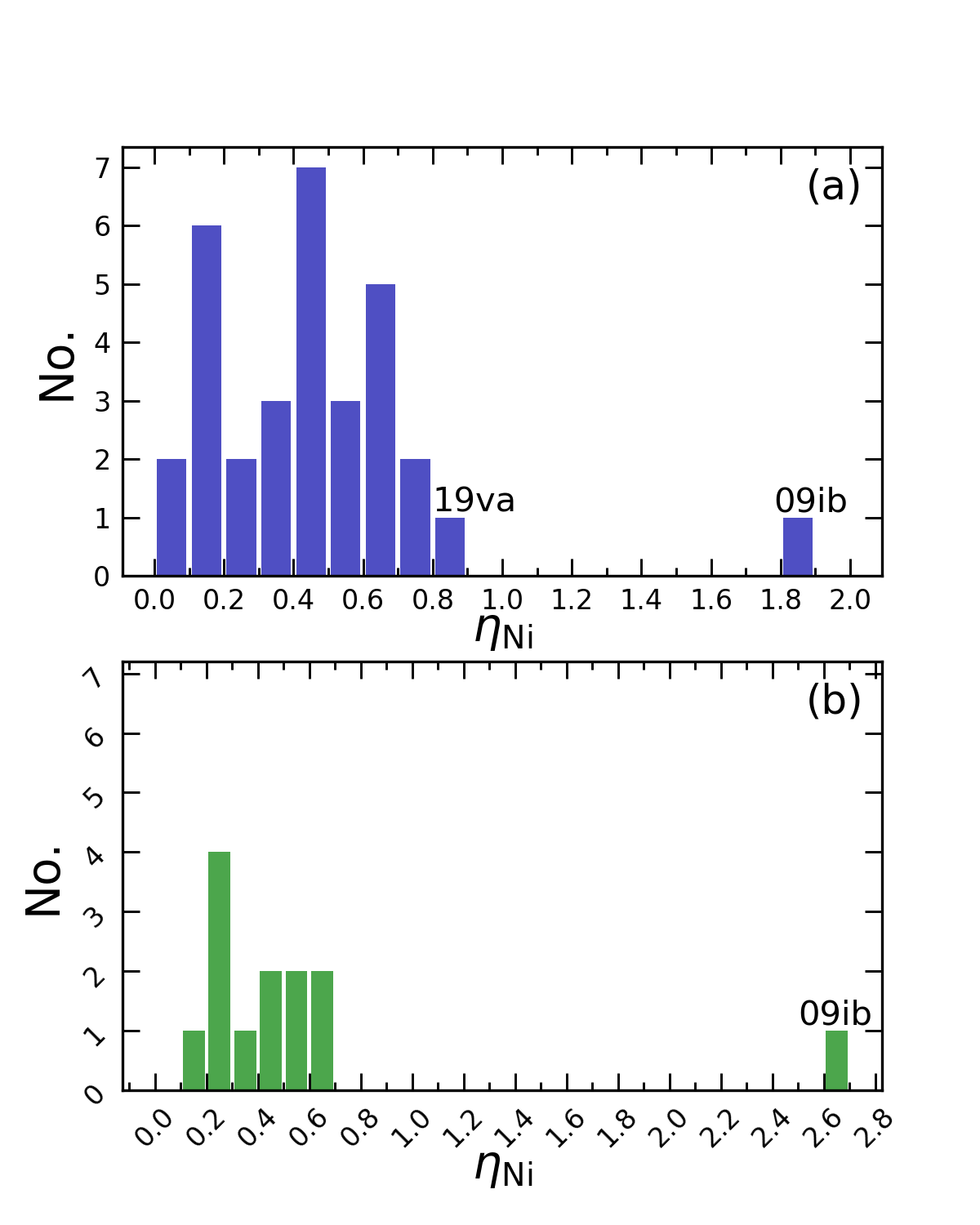}
    \caption{Histograms of $\eta_{\rm Ni}$. \textit{(a)}: $\eta_{\rm Ni}$ calculated from pseodo-bolometric light curves constructed by integrating luminosity over photometric data without IR bands. \textit{(b)}: $\eta_{\rm Ni}$ calculated from bolometric light curves constructed by data including IR bands. Locations of SN 2009ib and SN 2019va are marked.}
    \label{fig:hist_etaNi}
\end{figure}

Figure \ref{fig:hist_etaNi} shows the histograms of the $\eta_{\rm Ni}$ values listed in Table \ref{tab:etaNi}. We find that only SN 2009ib shows an $\eta_{\rm Ni}$ value larger than 1.0, which is possibly related to its extremely long plateau duration \citep{Takats2015}. Among the left sample, SN 2019va has the largest $\eta_{\rm Ni}$, with a value of 0.8. If taking into account the IR flux, an even larger value of $\eta_{\rm Ni}$ is expected for SN 2019va according to the above discussion. This indicates that $^{56}$Ni decay has a large influence on the plateau-phase light curve of SN 2019va.\\

\citet{Kozyreva2019} modelled the light curves of SNe II by including the influence of $^{56}$Ni decay during the plateau phase. In their simulations by adjusting the level of $^{56}$Ni mixing, they found the plateau-phase light curve can be affected (i.e., becoming brighter) from an earlier epoch when $^{56}$Ni is mixed into outer layers. However, \citet{Kozyreva2019} found they cannot determine the $^{56}$Ni-mixing level for most SNe II from their observed light curves because of the degeneracy in the simulated light curves. SN 2009ib is an outlier due to the extremely large $\eta_{\rm Ni}$ value, for which the simulation indicates that $^{56}$Ni is mixed throughout most of its hydrogen envelope. Although a relatively large $\eta_{\rm Ni}$ value is inferred for SN 2019va, we still cannot constrain the level of $^{56}$Ni mixing from the simulation results of \citet{Kozyreva2019}.\\

According to \citet{Kozyreva2019}, the effect of $^{56}$Ni decay on the plateau duration can be estimated by the following formula,
\begin{equation}
   \label{eq:plateauT}
   \frac{T_{pl}}{T_{pl}(\rm Ni = 0)} = (1+a\,\eta_{\rm Ni})^{1/6},
\end{equation}
where $T_{pl}$ is the observed plateau duration, $T_{pl}(\rm Ni = 0)$ is the expected plateau duration without the influence of $^{56}$Ni, and $a$ is a parameter with a value in the range of $2\sim6$, depending on the level of $^{56}$Ni mixing. Although the mixing level is uncertain for us, we know the plateau length of SN 2019va is extended at least by a factor of $(1+2\times\eta_{\rm Ni})^{1/6}=1.17$ (for $\eta_{\rm Ni}=0.8$) according to Equation (\ref{eq:plateauT}).\\

Note that SN 2005cs also displays a very flat plateau evolution (see Figure \ref{fig:compVpar}(a)), but its $\eta_{\rm Ni}$ value is much smaller than that of SN 2019va (see Table \ref{tab:etaNi}). This indicates that the physics underlying their very flat plateaus may not be the same. Thus, the influence of $^{56}$Ni decay on the plateau-phase light curve likely accounts partially for the scatter seen in the correlation between $M_{\rm max}^{V}$ and $s_2$ (see Figure \ref{fig:compVpar}(a) ).\\

\subsection{Explosion parameters}
\label{subsec:explpara}
To investigate the progenitor properties of SN 2019va, we use the empirical formulae proposed by \citet{Litvinova1985},
\begin{equation}
    \label{eq:LN1985}
    \begin{aligned}
       &\log_{10}(E_{\rm exp}) =  0.135M_V + 2.34\log_{10}(\Delta t) + 3.13\log_{10}(v_{\rm ph}) - 4.205, \\
       &\log_{10}(M_{\rm ej})  =  0.234M_V + 2.91\log_{10}(\Delta t) + 1.96\log_{10}(v_{\rm ph}) - 1.829, \\
       &\log_{10}(R_{\rm pSN}) = -0.572M_V - 1.07\log_{10}(\Delta t) - 2.74\log_{10}(v_{\rm ph}) - 3.350, \\
    \end{aligned}
\end{equation}
where $M_{V}$ is the $V$-band absolute magnitude at $t=50$\,d, $\Delta t$ (in units of days) is the plateau duration, $v_{\rm ph}$ (in units of $10^3$\,km\,s$^{-1}$) is the photospheric velocity at $t=50$\,d, $E_{\rm exp}$ (in units of $10^{51}$\,erg) is the explosion energy, $M_{\rm ej}$ (in units of solar masses) is the ejecta mass,  and $R_{\rm pSN}$ (in units of solar radii) is the pre-SN progenitor radius. \\

\begin{table}
	\centering
	\caption{Explosion parameters derived from the $V$-band light curve.}
	\label{tab:explpar}
	\begin{threeparttable}
	\begin{tabular}{ccccc} 
		\hline
		\hline
		Column No. & (1) & (2) & (3) & (4) \\
		\hline
		                                    & SN 2016gfy        &             &  SN 2019va   &       \\
		\hline
		$M_V$ (mag)                         & -16.74            &             & -16.72       &       \\
		\hline
		$\Delta t$ (days)                   & 90                & 79$^{*}$    & 100          &  85$^{*}$ \\
		\hline
		$v_{\rm ph}$ ($10^3$\,km\,s$^{-1}$) & 4.208             &             & 3.877        &       \\
		\hline
		\hline
		$E_{\rm exp}$ ($10^{51}$\,erg)      & 1.15              & 0.8         & 1.15         & 0.8  \\
		\hline
		$M_{\rm ej}$ ($M_{\odot}$)          & 14.6              & 10.0        & 17.1         & 10.8  \\
		\hline
		$R_{\rm pSN}$ ($R_{\odot}$)         & 270               & 310         & 289          & 342   \\
		\hline
		\hline
	\end{tabular}
	    \begin{tablenotes}
	        \footnotesize
	          \item[*]These columns list the new plateau durations that are corrected for the influence of $^{56}$Ni decay, as well as the recalculated explosion parameters using the new plateau durations (see text for more details).
	    \end{tablenotes}
	\end{threeparttable}
\end{table}

The calculated parameters of SN 2019va are listed in column (3) of Table \ref{tab:explpar}. Considering that SN 2016gfy shows many similarities with SN 2019va (plateau luminosity, weak brightening at the end of the plateau phase, photospheric velocity), we perform similar calculations for SN 2016gfy. The results are listed in column (1) of Table \ref{tab:explpar}. However, we find a not very reasonable result of a relatively large ejecta mass and a small progenitor radius. Actually, these limitations of the \citet{Litvinova1985} relations have been noticed and discussed by \citet{Hamuy2003} and \citet{Nadyozhin2003}. One main reason is that \citet{Litvinova1985} neglected the contribution of $^{56}$Ni decay to the plateau-phase light curve.\\

Since an extension factor of at least 1.17 has been inferred for the plateau duration of SN 2019va, thus we can recalculate the explosion parameters with a corrected plateau duration. The results are listed in column (4) of Table \ref{tab:explpar}. The same method is applied to SN 2016gfy, for which an extension factor of at least $(1+2\times\eta_{\rm Ni})^{1/6}=1.14$ (for $\eta_{\rm Ni}=0.6$) is estimated, and the recalculated results are listed in column (2) of Table \ref{tab:explpar}. As can be seen, the recalculated results are more reasonable for SNe II. In addition, we notice that the reestimated explosion parameters of SN 2016gfy and SN 2019va are comparable to each other, i.e., both having ejecta mass of $\sim10\,M_{\odot}$, progenitor radius of $\sim300\,R_{\odot}$, and explosion energy of $0.8\times10^{51}$\,erg. These similarities possibly indicate similar progenitors for SN 2016gfy and SN 2019va. Actually, one of the few observed differences between these two SNe II is that the plateau duration of SN 2019va is $\sim$10 days longer than that of SN 2016gfy. It is very likely that this difference comes from the influence of $^{56}$Ni decay, rather than the ejecta mass of their progenitors, especially considering that SN 2019va has a brighter tail and hence larger $^{56}$Ni mass (see Figure \ref{fig:compBolLC}). Although $^{56}$Ni decay can also make the plateau-phase light curve brighter, it is currently difficult to determine whether the magnitude at $t=50$\,d has been influenced without knowing the level of $^{56}$Ni mixing. This is further discussed in Appendix \ref{sec:flattenPlateau}.\\

\section{summary and conclusion}
\label{sec:conclusion}

We present extensive optical observations of SN 2019va, covering the rising phase of the light curve until the radioactive tail phase. The SN exploded at the edge of a spiral host galaxy, UGC 08577, which is inferred to have a relative low metallicity ($0.46\pm0.06$\,$Z_{\odot}$).\\

SN 2019va is found to have a very flat plateau lasting for $\sim100$\,days, with $\sim-16.72$\,mag in the $V$ band at $t=50$\,d. Moreover, it shows bluer $B-V$/$g-r$ colour evolution than the comparison SNe II. From the constructed pseudo-bolometric light curve, SN 2019va is found to show weak brightening towards the end of the plateau phase, which can be attributed to the influence of $^{56}$Ni decay on the plateau-phase light curve. \\

Extensive spectroscopic observations were also presented for SN 2019va. The metal lines, such as \ion{Na}{i} and \ion{Fe}{II}, are very weak in the plateau-phase spectra of SN 2019va, which is consistent with the low-metallicity environment of SN 2019va inferred from its host-galaxy spectrum.\\

We also estimate the $^{56}$Ni mass produced in SN 2019va and obtain a value of $0.088\pm0.018\,M_{\odot}$, which exceeds $\sim$90\% of the SN-II sample in \citet{Rodriguez2021}.\\

Following \citet{Nakar2016}, we calculate the $\eta_{\rm Ni}$ parameter of SN 2019va, and apply the calculations to an expanded SN-II sample. We find SN 2019va has the second largest $\eta_{\rm Ni}$ value among the sample, indicating that the plateau-phase light curve of SN 2019va is significantly influenced by $^{56}$Ni decay. After trying to remove this effect, we obtain a more reasonable estimate of the explosion/progenitor properties for SN 2019va. \\

Comparing with large samples of SNe II, we conclude that SN 2019va is somewhat peculiar in both photometric and spectroscopic evolution, with a flattening light curve and blue color during the plateau phase and weak metal lines in the spectra. SN 2016gfy is found to be another object showing these features. Given that both these two SNe II exploded in metal-poor environments, we propose that the appearances of these SNe II might be related to the properties of their progenitor system, e.g., metallicity. However, more objects like SN 2019va need to be discovered and quantitatively analyzed to test such a possibility. 

\section*{Acknowledgements}
The authors are grateful to some colleagues in the SN group or in THCA for useful suggestions on this paper. We thank Dr. C. Hao, A. Singh and J. L. Tonry for their help with our work in different aspects. The authors acknowledge support for observations from staff of XLT, LJT, APO. Funding for the LJT has been provided by the Chinese Academy of Sciences and the People's Government of Yunnan Province. The LJT is jointly operated and administrated by Yunnan Observatories and Center for Astronomical Mega-Science, CAS.\\
This work is supported by National Natural Science Foundation of China (NSFC grants 12033003, 11633002 and 11761141001) and the National Program on Key Research and Development Project (grant 2016YFA0400803). This work is partially supported by the Scholar Program of Beijing Academy of Science and Technology (DZ: BS202002) and the Tencent Xplorer Prize. We acknowledge the science research grants from the China Manned Space Project with No. CSM-CSST-2021-A12. A.P. Nagy is supported by the NKFIH/OTKA PD- 134434 grant. Y.-Z. Cai is funded by China Postdoctoral Science Foundation (grant no. 2021M691821). J.J.Z. is supported by the NSFC (grants 12173082,11773067), by the Youth Innovation Promotion Association of the CAS (grant 2018081), and by the Ten Thousand Talents Program of Yunnan for Top-notch Young Talents. T.M.Z. is supported by the NSFC (grant 11203034). \\

We acknowledge ESA Gaia, DPAC and the Photometric Science Alerts Team (http://gsaweb.ast.cam.ac.uk/alerts). This work has made use of data from the Asteroid Terrestrial-impact Last Alert System (ATLAS) project. The Asteroid Terrestrial-impact Last Alert System (ATLAS) project is primarily funded to search for near earth asteroids through NASA grants NN12AR55G, 80NSSC18K0284, and 80NSSC18K1575; byproducts of the NEO search include images and catalogs from the survey area. This work was partially funded by Kepler/K2 grant J1944/80NSSC19K0112 and HST GO-15889, and STFC grants ST/T000198/1 and ST/S006109/1. The ATLAS science products have been made possible through the contributions of the University of Hawaii Institute for Astronomy, the Queen’s University Belfast, the Space Telescope Science Institute, the South African Astronomical Observatory, and The Millennium Institute of Astrophysics (MAS), Chile.

\section*{Data Availability}
Photometric data of SN 2019va are presented in Table \ref{tab:TNTphot}, \ref{tab:ATophot}, \ref{tab:ATcphot} and \ref{tab:Gaiaphot}, and the spectroscopic data are available upon request.



\bibliographystyle{mnras}
\bibliography{SN2019va} 




\appendix
\section{Influence of $^{56}$Ni decay on the light curve at plateau phase: flatten the light curve}
\label{sec:flattenPlateau}
\citet{Nakar2016} defined two observables, 
\begin{equation}
    \label{eq:lambda_e}
    \Lambda_e = \frac{L_{25} \cdot (80\,\rm days)^2}{\int_0^{t_{\rm Ni}} t(L_{\rm bol}-Q_{\rm Ni}) \dif t},
\end{equation}
and
\begin{equation}
    \label{eq:lambda}
    \Lambda = \frac{L_{25} \cdot (80\,\rm days)^2}{\int_0^{t_{\rm Ni}} t\,L_{\rm bol} \dif t},
\end{equation}
where $L_{25}$ is the bolometric luminosity at $t=25$\,d, 80\,days is an empirical constant, $L_{\rm bol}$ is the constructed bolometric luminosity, $Q_{\rm Ni}$ is given in Equation (\ref{eq:Nidecay}), and $t_{\rm Ni}$ is the epoch separating the photospheric phase and the tail phase, which is defined in Equation (\ref{eq:etaNi}). \citet{Nakar2016} found that $2.5\log_{10}(\Lambda)$ is a good estimator of the magnitude decline between $t=25$\,d and $t=75$\,d for the bolometric light curve constructed from the observed data, and $2.5\log_{10}(\Lambda_e)$ is a good estimator of the magnitude decline between the above two epochs for the theoretical bolometric light curve if there was no contribution of $^{56}$Ni decay. Moreover, \citet{Nakar2016} argued that the influence of $^{56}$Ni decay on the luminosity at $t=25$\,d is very slight, thus $\Delta S = 2.5\log_{10}(\Lambda_e) - 2.5\log_{10}(\Lambda)$ [in units of mag\,(50\,d)$^{-1}$] represents how much the $^{56}$Ni decay flattens the plateau-phase light curve.\\

We set $A=\int_0^{t_{\rm Ni}} t Q_{\rm Ni} \dif t$, and $B=\int_0^{t_{\rm Ni}} t(L_{\rm bol}-Q_{\rm Ni}) \dif t$, so that we obtain $\eta_{\rm Ni}=\frac{A}{B}$ according to Equation (\ref{eq:etaNi}), $\Lambda_e = \frac{L_{25}\cdot(80\,\rm d)^2}{B}$ according to Equation (\ref{eq:lambda_e}), and $\Lambda = \frac{L_{25}\cdot(80\,\rm d)^2}{A+B}$ according to Equation (\ref{eq:lambda}). Then we derive
\begin{align}
    \label{eq:deltaS}
    \Delta S &= 2.5\log_{10}(\Lambda_e) - 2.5\log_{10}(\Lambda) \nonumber \\ 
    &=2.5\log_{10} \left[ \frac{L_{25}\cdot(80\,\rm d)^2}{B} \right] - 2.5\log_{10} \left[ \frac{L_{25}\cdot(80\,\rm d)^2}{A+B}\right] \nonumber \\
    &=2.5\log_{10}\left(\frac{1}{B}\right) - 2.5\log_{10}\left(\frac{1}{A+B}\right) \nonumber \\
    &= 2.5\log_{10}\left(\frac{A+B}{B}\right) \nonumber \\
    &= 2.5\log_{10}(\eta_{\rm Ni}+1).
\end{align}
For SN 2019va, $\Delta S$ can be calculated as 0.64\,mag\,(50\,d)$^{-1}$ (for $\eta_{\rm Ni}=0.80$), which means that, if there was no $^{56}$Ni, the light curve of SN 2019va would decline faster than the observed by about 0.64\,mag\,(50\,d)$^{-1}$ on average at the plateau phase (from $t=25$\,d to $t=75$\,d). For SN 2016gfy, the value of $\Delta S$ is calculated as 0.51\,mag\,(50\,d)$^{-1}$ (for $\eta_{\rm Ni}=0.60$). Note that Equation (\ref{eq:plateauT}) includes a parameter (i.e., $a$) describing the level of $^{56}$Ni mixing, while Equation (\ref{eq:deltaS}) lacks such a parameter. According to \citet{Kozyreva2019}, when $^{56}$Ni is mixed into the outer layers of the hydrogen envelope, the light curve will be brightened from an early epoch (e.g., before $t=50$\,d); if the $^{56}$Ni is confined to the inner layers of the SN ejecta, then the brightening will begin from a much later epoch (e.g., after $t=50$\,d). Therefore, it is difficult to accurately estimate whether and how much the light curve at $t=50$\,d is brightened due to the influence of $^{56}$Ni decay. However, adopting a simple assumption that the influence of $^{56}$Ni decay on the plateau slope is gradual with time, the magnitude difference at $t=50$\,d between the light curves with and without $^{56}$Ni can be roughly estimated by $\frac{\Delta S}{2}$. After correcting the $M_V$ values for this magnitude difference, and adopting the corrected plateau durations in Section \ref{subsec:explpara}, we recalculate the explosion parameters with Equation (\ref{eq:LN1985}) again. The recalculated results are listed in Table \ref{tab:explpar2}. The updated results still favour the conclusion we obtained in Section \ref{subsec:explpara}, that $^{56}$Ni decay can affect the plateau-phase light curve and hence affect the derived parameters describing the progenitor properties.
\begin{table}
	\centering
	\caption{Explosion parameters derived from the $V$-band light curve.}
	\label{tab:explpar2}
	\begin{threeparttable}
	\begin{tabular}{ccccc} 
		\hline
		\hline
		                                    & SN 2016gfy        &             &  SN 2019va   &       \\
		\hline
		$M_V$ (mag)                         & -16.74            & -16.48$^{*}$ & -16.72       &  -16.40$^{*}$ \\
		\hline
		$\Delta t$ (days)                   & 90                & 79$^{*}$    & 100          &  85$^{*}$ \\
		\hline
		$v_{\rm ph}$ ($10^3$\,km\,s$^{-1}$) & 4.208             &             & 3.877        &       \\
		\hline
		\hline
		$E_{\rm exp}$ ($10^{51}$\,erg)      & 1.15              & 0.9         & 1.15         & 0.9  \\
		\hline
		$M_{\rm ej}$ ($M_{\odot}$)          & 14.6              & 11.4        & 17.1         & 12.8  \\
		\hline
		$R_{\rm pSN}$ ($R_{\odot}$)         & 270               & 220         & 289          & 225   \\
		\hline
		\hline
	\end{tabular}
	    \begin{tablenotes}
	        \footnotesize
	          \item[*]These columns list the magnitudes and plateau durations corrected for the influences of $^{56}$Ni decay, as well as the recalculated explosion parameters (see Appendix \ref{sec:flattenPlateau} for more details).
	    \end{tablenotes}
	\end{threeparttable}
\end{table}

\section{Some tables}
\begin{table*}
	\centering
	\caption{Local reference stars in the field of SN 2019va.}
	\label{tab:refstars}
	\begin{threeparttable}
	\begin{tabular}{ccccccccc} 
		\hline
		Star & $RA$ (deg)  & $DEC$ (deg) & $B$ (mag) & $V$ (mag) & $g$ (mag) & $r$ (mag) & $i$ (mag) \\
		\hline
        1    & 203.888033  & 44.804294 & 12.547    & 11.975    & 12.507(002) & 11.895(001) & 11.806(002) \\
        2    & 203.721469  & 44.742893 & 15.955    & 15.068    & 15.445(003) & 14.794(003) & 14.571(003) \\
        3    & 203.838149  & 44.792061 & 15.680    & 15.128    & 15.311(003) & 14.914(003) & 14.771(003) \\
        4    & 203.756932  & 44.726836 & 16.981    & 16.162    & 16.705(004) & 15.827(003) & 15.521(004) \\
        5    & 203.734550  & 44.820268 & 17.411    & 16.458    & 16.797(004) & 16.178(004) & 15.914(004) \\
        6    & 203.928287  & 44.784795 &	--        & --        & 17.053(004) & 16.143(004) & 15.770(003) \\
        7    & 203.931533  & 44.781422 & --        & --        & 16.930(004) & 16.517(004) & 16.368(004) \\
        8    & 203.846728  & 44.721422 & --        & --        & 17.297(005) & 16.895(004) & 16.744(005) \\
        9    & 203.925544  & 44.731447 & --        & --        & 17.396(006) & 16.469(004) & 16.050(004) \\
        10   & 203.867264  & 44.734644 & --        & --        & 17.498(005) & 16.635(004) & 16.312(004) \\
        11   & 203.870500  & 44.738711 & --        & --        & 17.592(005) & 17.297(005) & 17.182(005) \\
		\hline
	\end{tabular}
	    \begin{tablenotes}
	        \footnotesize
	          \item[] See Figure \ref{fig:image} for the star numbers. $BV$ bands are in Vega magnitude system, and $gri$ bands are in AB magnitude system. Uncertainties in parentheses are in units of 0.001 mag.
	    \end{tablenotes}
	\end{threeparttable}
\end{table*}

\begin{table*}
	\centering
	\caption{Photometry of SN 2019va from Tsinghua-NAOC 0.8m telescope.}
	\label{tab:TNTphot}
	\begin{threeparttable}
	\begin{tabular}{cccccc} 
		\hline
		MJD      & $B$(mag)    & $V$(mag)    & $g$(mag)    & $r$(mag)    & $i$(mag)   \\
		\hline
        58504.92 & 16.531(036) & 16.502(023) & 16.448(021) & 16.540(027) & 16.680(037) \\ 
        58506.90 & 16.611(048) & 16.558(030) & 16.410(026) & 16.567(030) & 16.730(042) \\ 
        58507.89 & 16.598(038) & 16.540(025) & 16.463(021) & 16.494(028) & 16.652(036) \\ 
        58515.89 & 16.724(032) & 16.512(021) & 16.519(019) & 16.410(027) & 16.524(035) \\ 
        58533.92 & 17.050(078) & 16.501(052) & 16.789(043) & 16.398(042) & 16.396(055) \\ 
        58534.81 & 17.141(031) & 16.547(022) & 16.777(024) & 16.413(029) & 16.488(045) \\ 
        58537.69 & 17.258(073) & 16.514(032) & 16.780(024) & 16.438(031) & 16.417(048) \\ 
        58538.82 & 17.304(077) & 16.521(037) & 16.841(024) & 16.389(035) & 16.482(050) \\ 
        58539.68 & 17.304(037) & 16.559(026) & 16.790(024) & 16.426(031) & 16.455(047) \\ 
        58545.71 & 17.323(031) & 16.529(022) & 16.803(024) &     --      &     --      \\ 
        58548.85 & 17.366(035) & 16.531(021) & 16.848(024) & 16.379(034) & 16.417(045) \\ 
        58562.73 &     --      &     --      & 16.950(049) & 16.295(033) & 16.340(042) \\ 
        58563.73 & 17.406(068) & 16.520(041) & 16.850(045) & 16.356(052) & 16.416(079) \\ 
        58566.68 & 17.407(031) & 16.481(022) & 16.826(024) & 16.287(027) & 16.335(044) \\ 
        58580.71 & 17.424(063) & 16.500(024) & 16.850(020) & 16.284(027) & 16.348(036) \\ 
        58582.80 & 17.450(046) & 16.519(023) & 16.869(021) & 16.293(027) & 16.325(035) \\ 
        58589.73 &     --      &     --      & 16.816(106) & 16.534(122) & 16.381(133) \\ 
        58590.79 &     --      &     --      & 16.915(096) & 16.279(098) &     --      \\ 
        58594.69 & 17.710(101) & 16.534(047) & 16.995(044) & 16.355(037) & 16.370(053) \\ 
        58603.67 & 17.925(033) & 16.746(023) & 17.210(024) & 16.493(033) & 16.493(044) \\ 
        58604.77 & 17.932(034) & 16.774(022) & 17.248(024) & 16.506(030) & 16.526(044) \\ 
        58605.77 & 17.996(033) & 16.813(022) & 17.276(024) & 16.536(027) & 16.531(044) \\ 
        58606.74 & 18.102(034) & 16.842(023) & 17.344(024) & 16.549(028) & 16.551(044) \\ 
        58608.70 & 18.177(041) & 16.922(022) & 17.429(021) & 16.620(027) & 16.596(035) \\ 
        58612.70 & 18.481(035) & 17.133(022) & 17.672(024) & 16.752(027) & 16.694(044) \\ 
        58613.57 & 18.591(040) & 17.151(024) & 17.758(025) & 16.766(028) & 16.710(044) \\ 
        58623.66 & 19.375(056) & 17.823(028) & 18.549(030) & 17.410(028) & 17.271(044) \\ 
        58630.66 & 19.948(177) & 18.229(040) & 18.947(038) & 17.824(030) & 17.648(038) \\ 
        58633.61 & 20.110(127) & 18.301(045) & 19.126(058) & 17.938(031) & 17.726(041) \\ 
        58645.69 &     --      &     --      &     --      & 18.036(059) & 17.966(074) \\ 
        58648.69 &     --      & 18.518(165) & 19.290(053) & 18.143(038) & 17.902(080) \\ 
        58663.64 & 20.412(181) & 18.646(053) & 19.419(039) & 18.244(038) & 18.223(044) \\ 
        58664.64 &     --      &     --      & 19.415(044) & 18.259(033) & 18.126(047) \\ 
		\hline
	\end{tabular}
        \begin{tablenotes}
             \footnotesize
             \item Note: numbers in bracket are uncertainties in units of 0.001 mag.
        \end{tablenotes}	\end{threeparttable}
\end{table*}

\begin{table*}
	\centering
	\caption{$o$-band photometry of SN 2019va from ATLAS.}
	\label{tab:ATophot}
	\begin{threeparttable}
	\begin{tabular}{cccccccc} 
		\hline
		MJD      & mag         & MJD      & mag         & MJD      & mag         & MJD      & mag         \\
		\hline
        58494.67 & >18.50    & 58568.52 & 16.23(01) & 58622.37 & 17.09(03) & 58682.32 & 17.99(07) \\
        58498.67 & 16.91(03) & 58570.54 & 16.23(01) & 58624.31 & 17.24(01) & 58684.28 & 18.19(03) \\
        58500.67 & 16.72(02) & 58572.51 & 16.21(01) & 58626.37 & 17.35(02) & 58688.27 & 18.11(03) \\
        58502.61 & 16.61(01) & 58574.57 & 16.23(01) & 58630.35 & 17.53(02) & 58690.28 & 18.20(04) \\
        58506.66 & 16.48(02) & 58578.54 & 16.22(01) & 58634.36 & 17.64(02) & 58716.25 & 18.32(05) \\
        58508.64 & 16.42(01) & 58582.49 & 16.24(01) & 58642.28 & 17.71(02) & 58722.24 & 18.47(05) \\
        58510.64 & 16.43(01) & 58588.52 & 16.26(01) & 58644.33 & 17.70(02) & 58832.63 & 19.47(15) \\
        58514.63 & 16.39(01) & 58590.51 & 16.25(02) & 58646.33 & 17.76(04) & 58836.64 & 19.80(17) \\
        58518.60 & 16.34(01) & 58594.47 & 16.29(02) & 58648.32 & 17.77(03) & 58846.58 & 19.77(11) \\
        58522.65 & 16.34(01) & 58596.49 & 16.31(01) & 58650.35 & 17.73(05) & 58850.61 & 19.71(11) \\
        58530.59 & 16.34(01) & 58598.47 & 16.32(00) & 58654.30 & 17.86(03) & 58866.60 & 19.68(10) \\
        58538.60 & 16.32(01) & 58600.49 & 16.33(01) & 58656.33 & 17.87(03) & 58868.62 & 19.98(13) \\
        58542.60 & 16.33(01) & 58606.49 & 16.44(01) & 58658.32 & 17.85(05) & 58870.55 & 20.02(15) \\
        58546.60 & 16.32(01) & 58610.47 & 16.55(01) & 58660.30 & 17.79(04) & 58874.61 & 20.34(18) \\
        58550.57 & 16.30(01) & 58614.39 & 16.70(01) & 58672.29 & 17.94(03) & 58882.56 & 20.22(16) \\
        58554.54 & 16.28(01) & 58616.39 & 16.79(01) & 58675.30 & 17.96(05) & 58902.56 & 20.22(17) \\
        58558.56 & 16.27(01) & 58618.38 & 16.94(02) & 58678.29 & 18.04(05) & 58906.50 & 20.09(10) \\
        58562.49 & 16.26(02) & 58620.37 & 17.02(03) & 58680.28 & 18.08(05) & 58910.51 & 20.38(14) \\
		\hline
	\end{tabular}
        \begin{tablenotes}
             \footnotesize
             \item Note: numbers in bracket are uncertainties in units of 0.01 mag.
        \end{tablenotes}	\end{threeparttable}
\end{table*}

\begin{table*}
	\centering
	\caption{$c$-band photometry of SN 2019va from ATLAS.}
	\label{tab:ATcphot}
	\begin{threeparttable}
	\begin{tabular}{cccccccc} 
		\hline
		MJD      & mag         & MJD         & mag      & MJD      & mag         & MJD      & mag         \\
		\hline
        58520.66 & 16.51(01) & 58604.48 & 16.83(01) & 58664.29 & 18.69(04) & 58696.26 & 19.02(10) \\
        58576.55 & 16.58(01) & 58612.39 & 17.14(01) & 58668.35 & 18.62(04) & 58700.29 & 19.02(07) \\
        58580.53 & 16.57(01) & 58636.40 & 18.38(02) & 58692.28 & 19.00(05) & 58724.24 & 19.30(14) \\
		\hline
	\end{tabular}
        \begin{tablenotes}
             \footnotesize
             \item Note: numbers in bracket are uncertainties in units of 0.01 mag.
        \end{tablenotes}	\end{threeparttable}
\end{table*}

\begin{table*}
	\centering
	\caption{$G$-band photometry of SN 2019va from Gaia.}
	\label{tab:Gaiaphot}
	\begin{threeparttable}
	\begin{tabular}{cccccccc} 
		\hline
		MJD      & mag         & MJD         & mag      & MJD      & mag         & MJD      & mag         \\
		\hline
        58507.43 & 16.44(02) & 58621.53 & 17.37(00) & 58745.48 & 19.03(01) & 58876.73 & 20.53(00) \\
        58549.81 & 16.46(01) & 58692.14 & 18.51(00) & 58790.06 & 19.46(00) & 58899.33 & 20.86(00) \\
        58558.29 & 16.46(01) & 58736.79 & 18.92(01) & 58807.84 & 19.72(00) &          &           \\
		\hline
	\end{tabular}
        \begin{tablenotes}
             \footnotesize
             \item Note: numbers in bracket are uncertainties in units of 0.01 mag.
        \end{tablenotes}	\end{threeparttable}
\end{table*}

\begin{table*}
	\centering
	\caption{Journal of spectroscopic observations of SN 2019va.}
	\label{tab:spectra}
	\begin{threeparttable}
	\begin{tabular}{lcccclc} 
		\hline
		No. & UT Date        & MJD     & Epoch (day)$^{**}$ & Exp. (s) & Telescope+Inst. & Range (\AA) \\
		\hline
		1   & Jan. 18th 2019 & 58501.9 & +5.2            & 1500     & LJT+YFOSC       & 3500-8600   \\
        2   & Jan. 29th 2019 & 58512.8 & +16.1           & 3600     & XLT+BFOSC       & 3900-8600   \\
        3   & Feb. ~1st 2019 & 58515.9 & +19.2           & 3600     & XLT+BFOSC       & 4400-8500   \\
        4   & Feb. ~8th 2019 & 58522.8 & +26.1           & 3000     & XLT+BFOSC       & 4400-8500   \\
        5   & Feb. 15th 2019 & 58529.7 & +33.0           & 1350     & LJT+YFOSC       & 3500-8600   \\
        6$^{*}$   & Mar. 14th 2019 & 58556.3 & +59.6           & 1200     & P60+SEDM        & 3800-9000   \\
        7   & Mar. 25th 2019 & 58567.9 & +71.2           & 2100     & XLT+BFOSC       & 4900-8000   \\
        8$^{*}$   & Mar. 28th 2019 & 58570.3 & +73.6           & 1134     & APO+DIS         & 3500-9700   \\
        9   & May~ ~6th 2019 & 58609.6 & +112.9          & 2700     & XLT+BFOSC       & 4100-8700   \\
        10  & May~ ~7th 2019 & 58611.4 & +114.7          & 2400     & APO+DIS         & 3500-9000   \\
        11  & May~ 31st 2019 & 58634.8 & +138.1          & 1700     & LJT+YFOSC       & 3500-8600   \\
		\hline
	\end{tabular}
	    \begin{tablenotes}
	        \footnotesize
	          \item[*] These two spectra are taken from the TNS.
	          \item[**] The epoch is relative to the explosion date, MJD=58496.7.
	    \end{tablenotes}
	\end{threeparttable}
\end{table*}

\begin{table*}
	\centering
	\caption{Information of Supernovae used in the paper.}
	\label{tab:info}
	\begin{threeparttable}
	\begin{tabular}{ccccc} 
		\hline
		SN name     & distance modulus & explosion epoch      & $E(B-V)_{\rm total}$ & reference \\
		            & (mag)            & (JD)                 & (mag)                &           \\
		\hline
		SN 1987A     & 18.49$\pm$0.13         & 2446859.82           & 0.175              & \citet{Woosley1987, Hamuy1988, Storm2004, Taddia2012}       \\
		SN 1999em    & 30.34$\pm$0.07   & 2451475.00$\pm$1.0   & 0.10 & \citet{Leonard2002, Leonard2003, Elmhamdi2003}       \\
        SN 2005cs    & 29.26$\pm$0.33   & 2453549.00$\pm$0.5   & 0.05 & \citet{Pastorello2006, Pastorello2009}       \\
        SN 2009ib    & 31.48$\pm$0.31   & 2455041.80$\pm$2.0   & 0.16 &\citet{Takats2015, Valenti2016}       \\
        SN 2013by    & 30.81$\pm$0.15   & 2456404.00$\pm$2.0   & 0.23 & \citet{Valenti2015, Valenti2016} \\
        SN 2016gfy   & 32.36$\pm$0.18   & 2457641.40$\pm$0.9   & 0.21 & \citet{Singh2019} \\
        SN 2017eaw   & 29.44$\pm$0.21   & 2457885.70$\pm$0.1   & 0.30 & \citet{Rui2019, VanDyk2019} \\
        SN 2019va    & 33.20$\pm$0.15   & 2458497.20$\pm$2.0   & 0.016 & This paper \\
        \hline
        SN 2001X     & 31.59$\pm$0.11   & 2451963.00$\pm$5.0   & 0.11  & \citet{Faran2014a, Valenti2016}  \\           
        SN 2009kr    & 32.09$\pm$0.15   & 2455140.50$\pm$2.0   & 0.07  & \citet{Elias_Rosa2010, Valenti2016}     \\
		SN 2013bu    & 30.79$\pm$0.08   & 2456399.80$\pm$4.5   & 0.08  & \citet{Valenti2016} \\
        SN 2013fs    & 33.45$\pm$0.15   & 2456571.12$\pm$0.5   & 0.04  & \citet{Valenti2016} \\
        LSQ13dpa    & 35.08$\pm$0.15   & 2456642.70$\pm$2.0   & 0.04  & \citet{Valenti2016} \\
        SN 2014cy    & 31.87$\pm$0.15   & 2456900.00$\pm$1.0   & 0.05  & \citet{Valenti2016, Dastidar2021} \\    
        SN 2014dw    & 32.46$\pm$0.15   & 2456958.00$\pm$10.0  & 0.22  & \citet{Valenti2016} \\
        SN 2014G     & 31.90$\pm$0.15   & 2456668.35$\pm$1.0   & 0.21  & \citet{Terreran2016, Valenti2016} \\
        ASASSN14dq  & 33.26$\pm$0.15   & 2456841.50$\pm$5.5   & 0.07  & \citet{Valenti2016} \\
        ASASSN14gm  & 31.74$\pm$0.15   & 2456901.00$\pm$1.5   & 0.10  & \citet{Valenti2016} \\
        ASASSN14ha  & 29.53$\pm$0.50   & 2456910.50$\pm$1.5   & 0.01  & \citet{Valenti2016} \\
        SN 2015W     & 33.74$\pm$0.15   & 2457025.00$\pm$10.0  & 0.15  & \citet{Valenti2016} \\
		\hline
	\end{tabular}
	    \begin{tablenotes}
	        \footnotesize
	          \item[] 
	          \item[]
	    \end{tablenotes}
	\end{threeparttable}
\end{table*}

\bsp	
\label{lastpage}
\end{document}